\documentclass{aa}  

\usepackage{acronym}
\usepackage{color, colortbl}
\usepackage{graphicx}
\usepackage[hidelinks]{hyperref} 
\usepackage{lscape}
\usepackage{lipsum}
\usepackage{placeins}
\usepackage{subcaption}         
\usepackage{txfonts}
\usepackage[dvipsnames]{xcolor}

\newcommand{\refsec}[1]{Sect.~\ref{#1}}
\newcommand{\reftab}[1]{Table~\ref{#1}}
\newcommand{\refeq}[1]{Eq.~(\ref{#1})}
\newcommand{\reffig}[1]{Fig.~\ref{#1}}

\newacro{BH}{black hole}
\newacro{BNS}{binary neutron star}
\newacro{EOS}{equation of state}
\newacro{NS}{neutron star}
\newacro{NSE}{nuclear statistical equilibrium}

\begin{document}

   \title{The \texttt{MARTINI} platform\\
   II. Tracing $r$-process nucleosynthesis in neutron star mergers\\ with long-lived remnants}

   \author{E. Loffredo\inst{1,2}\fnmsep\thanks{Corresponding author: eleonora.loffredo@inaf.it}
   \and D. Vescovi\inst{2,3}
   \and A. Perego\inst{4,5}
   \and F. M. Guercilena\inst{4,5}
   \and S. Cristallo\inst{2,3}
    }

   \institute{INAF – Osservatorio Astronomico di Roma, Via Frascati 33, I-00078 Monte Porzio Catone (Rome), Italy
   \and INAF – Osservatorio Astronomico d’Abruzzo, via Mentore Maggini snc, 64100 Teramo, Italy
   \and INFN, Sezione di Perugia, Via A. Pascoli snc, 06123, Perugia, Italy
   \and Dipartimento di Fisica, Università di Trento, Via Sommarive 14, 38123 Trento, Italy 
   \and INFN-TIFPA, Trento Institute for Fundamental Physics and Applications, Via Sommarive 14, I-38123 Trento, Italy}

   \date{Received January 30, 20XX}

  \abstract
  {Binary neutron star (BNS) mergers are among the primary sites of the rapid neutron-capture ($r$-) process. In mergers with long-lived remnants, sustained neutrino irradiation raises the electron fraction of the disc wind ejecta and modifies the resulting nucleosynthesis pattern. Yields from numerical-relativity simulations with accurate neutrino transport are therefore needed to quantify the contribution of BNS mergers to the Galactic production of heavy elements and to interpret kilonova spectra.}
  {We present and publicly release a set of tracer trajectories and nucleosynthesis yields for BNS mergers that form long-lived neutron star remnants, obtained from full general-relativistic simulations with M1 neutrino transport.}
  {We performed eight numerical-relativity simulations of BNS mergers with the M1 neutrino transport scheme, spanning two equations of state (DD2 and SFHo), different total binary masses, and mass ratios. All simulations produced long-lived neutron star remnants. We recorded the thermodynamic properties of the dynamical and wind ejecta crossing a detector surface and the thermodynamic evolution of passive tracer particles. The tracers were post-processed with the nuclear reaction network \texttt{WinNet} to compute the nucleosynthesis yields up to $t=1$ Gyr.}
  {We publicly release, through the \texttt{MARTINI} platform, the nucleosynthesis yields at $1$ day and $1$ Gyr and the tracer trajectories of all simulations. At $1$ day, the production of light and first-peak elements (He, Ni, Sr, and Y) is similar across different BNS merger models and is primarily enhanced by the neutrino-irradiated disc wind. The yields of heavy nuclei ($Z \geq 52$) vary by more than one order of magnitude, with unequal-mass binaries generally producing the largest amounts owing to their larger fraction of neutron-rich ejecta. Our He yields, particularly in the wind, are larger than those of recent studies based on an M0 neutrino scheme, consistent with the stronger neutrino irradiation resulting from an M1 scheme. At $1$ Gyr, all models reproduce the second and third solar $r$-process peaks, but differ in the first peak, the lanthanide region, the mass range between the first and second peaks, and the trans-lead region, reflecting the differences in the electron fraction distribution of the ejecta and the relative contributions of the dynamical and wind components. Two equal-mass models provide a close match to the solar pattern overall.}
  {}

   \keywords{dense matter --
                equation of state --
                nuclear reactions, nucleosynthesis, abundances
               }

   \maketitle
   \nolinenumbers

\section{Introduction}

The astrophysical origin of elements heavier than iron (Fe) remains one of the longest-standing problems in nuclear astrophysics \citep{NAP10079}. Approximately half of these nuclei are synthesised through the rapid neutron-capture ($r$-) process \citep{Burbidge1957RvMP}, which is responsible for the production of the heaviest stable elements in nature, including the actinides $^{232}$Th, $^{235}$U, and $^{238}$U \citep[for recent reviews on $r$-process, see][]{Thielemann2017ARNPS, Cowan2021RvMP}. Because the $r$-process requires extremely neutron-rich conditions, with neutron densities exceeding $\sim10^{26}~{\rm cm^{-3}}$ to produce elements up to actinides, and a short timescale ($\sim 1~ 
{\rm s}$), identifying the astrophysical environments in which it operates has been a major challenge for decades. 
Several astrophysical sites have been proposed as potential sources of the $r$-process, including core-collapse supernovae, magneto-rotational explosions, collapsars, and compact-object mergers. While some of these environments may contribute to a weak $r$-process \citep{Wanajo2006NuPhA, Honda2006ApJ}, growing theoretical and observational evidence indicates that \ac{BNS} mergers, and possibly \ac{NS}-\ac{BH} mergers, are among the primary production sites of the main $r$-process responsible for the Solar abundance pattern \citep[e.g.][]{Sneden2008ARA&A, Rosswog2018A&A, Horowitz2019AnPhy, Shibata2019ARNPS}.

Compelling evidence for this scenario came with the epochal discovery of GW170817, the first gravitational-wave detection of a \ac{BNS} merger, and the observation of its associated UV/optical/NIR counterpart, the kilonova AT2017gfo \citep{Abbott2017PhRvLgw170817, Abbott2017ApJgw170817}. AT2017gfo exhibited clear signatures of freshly synthesised $r$-process material, providing the first direct observational confirmation that \ac{BNS} mergers produce heavy elements through $r$-process \citep[e.g.][]{Pian2017Natur, Smartt2017Natur}.
Additional observational evidence has further strengthened interest in understanding $r$-process nucleosynthesis in the ejecta of compact binary mergers. In particular, the abundance patterns observed in extremely metal-poor stars, including the so-called actinide-boost stars characterised by enhanced thorium (Th) and uranium (U) abundances relative to lighter $r$-process elements such as europium (Eu), provide a unique probe of the astrophysical conditions under which the heaviest nuclei are synthesised and raise important questions about the diversity of $r$-process sites \citep[e.g.][]{Ji2018ApJ,Holmbeck2018ApJ,Holmbeck2019ApJ,Eichler2019ApJ,Kullmann2023MNRAS}. \citet{Eichler2019ApJ} showed that the Th/Eu ratio is highly sensitive to both the thermodynamic conditions of the ejecta and the adopted nuclear input, particularly the nuclear mass model, suggesting that the observed diversity of Th/Eu ratios may arise either from different ejecta components within \ac{BNS} mergers or from multiple astrophysical $r$-process sites.

In this context, accurate nucleosynthesis predictions from \ac{BNS} mergers are essential not only to quantify the contribution of these events to the Galactic production of heavy elements, but also to interpret kilonova observations and to assess whether \ac{BNS} mergers can reproduce the chemical signatures observed in metal-poor and actinide-boost stars.

The nucleosynthesis outcome from a \ac{BNS} merger depends strongly on the properties of the ejected matter, in particular its electron fraction, entropy, and expansion velocity \citep{Hoffman1997NuPhA, Hoffman1997ApJ}. 
These properties depend sensitively on several intrinsic binary properties, such as the total mass, mass ratio and spins. Additionally, they vary subject to several microphysics inputs, including the nuclear \ac{EOS} and the treatment of weak interactions, requiring fully general-relativistic simulations coupled to neutrino-radiation hydrodynamics (see \citealt{Radice2018ApJ, Radice2020ARNPS, Perego2021hgwa} for recent reviews).
In particular, the lifetime of the merger remnant plays a crucial role in shaping the properties of the remnant accretion disc and the subsequent outflows. Long-lived \ac{NS} remnants sustain neutrino irradiation over timescales comparable to the disc lifetime, increasing the electron fraction of disc wind ejecta and substantially modifying the resulting nucleosynthesis pattern \citep[e.g.][]{Fujibayashi2020ApJ, Kawaguchi2021ApJ, Just2023ApJ, Bernuzzi2025MNRAS,Sneppen2026PhRvD}. 
The predicted ejecta composition is also highly sensitive to the adopted neutrino transport scheme. Specifically, the M1 neutrino transport scheme \citep{Foucart2016PRDa,Radice2022MNRAS,Zappa2023MNRAS} consistently predicts broader electron fraction distributions in the neutrino-driven wind than the simpler M0 approximation, leading to enhanced production of light and Fe-peak nuclei \citep[e.g.][]{Foucart2016PRDa, Vincent2020PhRvD, Radice2022MNRAS, Radice2023ApJ, Radice2024JPhCS, Bernuzzi2025MNRAS, Jacobi2026ApJ, Sneppen2026PhRvD}. The angular structure of the neutrino irradiation in M1 simulations and its imprint on the electron fraction was recently quantified by \citet{Cokluk2026arXiv260816275C} for the set of \ac{BNS} merger simulations considered in this work. They found that the quadrupole-dominated neutrino radiation field drives the polar material close to the equilibrium electron fraction set by neutrino absorption, with values of about 0.50--0.55 at late times, whereas the equatorial material remains more neutron-rich and below its equilibrium value. The anisotropy of the neutrino flux therefore leaves a robust polar–equatorial imprint on the composition of the ejecta, which is expected to affect both the nucleosynthesis and the kilonova emission.

Among recent studies, \citet{Ricigliano2024MNRAS2} investigated the impact of the nuclear \ac{EOS} on the nucleosynthesis and kilonova emission from \ac{BNS} mergers using tracer particles extracted from simulations employing an M0 neutrino transport scheme. They showed that both the isotopic yields and the kilonova emission depend sensitively on the \ac{EOS}, highlighting significant degeneracies between merger dynamics and nuclear microphysics.
More recently, the production of helium (He) and other light elements in neutrino-irradiated winds from mergers with long-lived remnants has gained significant attention because of their potential signatures in kilonova spectra \citep{Perego2022ApJ, Sneppen2024A&A, Sneppen2026PhRvD, Arya2026arXiv, Chiba2026arXiv}. In particular, He abundance has been proposed as a diagnostic of the remnant lifetime and, consequently, of the nuclear \ac{EOS}, since long-lived remnants are expected to synthesize substantially more He than promptly collapsing systems \citep{Sneppen2026PhRvD}.
More generally, these developments highlight the need for nucleosynthesis calculations based on numerical-relativity simulations with accurate modelling of neutrino irradiation, and that provide detailed isotopic yields for both the dynamical and disc wind ejecta. This need is reinforced by recent long-duration simulations, which demonstrated that late-time disc outflows can significantly modify the final abundance pattern compared with simulations terminated shortly after merger, emphasizing the importance of modelling the long-term evolution of the remnant and its winds \citep[e.g.][]{Fujibayashi2018ApJ, Fujibayashi2020ApJ, Kiuchi2023PhRvL, Fahlman2023MNRAS, Sprouse2023ApJ, Radice2023ApJ, LongoMicchi2026arXiv}.

In this paper, we present publicly available $r$-process nucleosynthesis yields from \ac{BNS} mergers with long-lived remnants, obtained for the first time by post-processing tracer particles extracted from simulations in full general relativity employing the M1 neutrino transport scheme. Our sample spans two nuclear \ac{EOS}, a range of binary total masses, and different mass ratios enabling us to investigate how the properties of the binary influence the resulting nucleosynthesis. The neutrino emission and the electron fraction evolution of these simulations were analysed in detail in \citet{Cokluk2026arXiv260816275C}. In this work, we focus on the nucleosynthesis. Elemental and isotopic yields for each simulation are obtained by processing tracer particles with the \texttt{WinNet} nuclear network \citep{Reichert2023ApJS}.
Since neutrino and viscosity driven winds are expected to unbind up to $\sim 40\%$ of the remnant accretion disc mass \citep[e.g.][]{Siegel2017PhRvL, Fujibayashi2018ApJ, Fujibayashi2020ApJ, Fahlman2023MNRAS}, we provide nucleosynthesis yields for multiple prescriptions of the wind ejecta mass to account for the current uncertainties associated with the late-time disc outflows. The complete set of isotopic and elemental yields, together with the corresponding tracer trajectories, is publicly released through the \texttt{MARTINI} platform \citep{Bezmalinovich2026A&A}, providing a new resource for kilonova modelling, Galactic chemical evolution studies, and comparisons with the chemical abundance patterns observed in metal-poor and actinide-boost stars.

The paper is structured as follows. In \refsec{sec:method}, we describe the numerical relativity simulations, the nucleosynthesis calculations, and the methodology followed to compute the nucleosynthesis yields. In \refsec{sec:results}, we present the properties of the merger ejecta, the evolution of the tracer particles, and the resulting yields at 1 day and 1 Gyr after the merger. In \refsec{sec:discussion}, we compare our results to several studies in the literature. In \refsec{sec:Martini}, we describe the implementation of the computed yields within the \texttt{MARTINI} platform and provide an example of data use. Finally, \refsec{sec:conclusions} summarises our main results and presents our conclusions.

\section{Method}\label{sec:method}

\subsection{BNS merger simulations}

We perform eight \ac{BNS} merger simulations in full general relativity, spanning different component masses and two nuclear \acp{EOS}. The key properties of each simulation are summarised in \reftab{tab:bns_sims}, where $M_{1,2}$ denote the gravitational masses of the two \acp{NS} measured at infinity, $q = M_1/M_2 \leq 1$ is the binary mass ratio, $M_{\rm tot}$ is the total mass of the binary,  $t_{\rm mrg}$ is the merger time, and $t_{\rm end}$ is the final simulation time. In all cases, the merger results in a long-lived \ac{NS} remnant (i.e. a \ac{NS} surviving for at least 50 ms -- 80 ms after the merger), surrounded by a matter disc.

Each simulation is initialised using the \texttt{Lorene} library \citep{Gourgoulhon2001PhRvD, Gourgoulhon2016} with two irrotational \acp{NS} in quasi-circular orbit placed at a separation of $40-50~$km, corresponding to a few orbits before merger.  The two \acp{NS} are in neutrino-less beta and hydrostationary equilibrium. The initial configuration is then evolved by solving Einstein's equations coupled to general relativistic hydrodynamics in a $3+1$ dimension spacetime. The evolution of spacetime is performed through the \texttt{CTGamma} code \citep{Reisswig2013PhRvL}, which is part of \texttt{Einstein Toolkit} \citep{Loffler2012CQGra, Etienne2021zndo}, employing the Z4c formulation of Einstein’s equations \citep{Bernuzzi2010PrD, Hilditch2013PrD}, the 1+log condition for the lapse and the integrated Gamma-driver condition for the shift. Fourth-order finite-difference operators are used to evolve the metric fields, applying the Kreiss–Oliger dissipation to ensure the nonlinear stability of the evolution. The general relativistic hydrodynamics simulation of matter evolution is performed through the \texttt{WhiskyTHC} code \citep{Radice2012AAP, Radice2014APS, Radice2014MNRAS, Radice2014CQG} adopting flux-conservative high-resolution shock-capturing schemes. 
We use fifth-order monotonicity-preserving reconstruction \citep{Suresh1997JCoPh} in combination with an HLLE approximate Riemann solver \citep{Harten1983SIAM, Einfeldt1988SJNA}. The hydrodynamic evolution and spacetime dynamics are coupled through the method of lines adopting a third order Runge–Kutta scheme for time integration with a fixed Courant–Friedrichs–Lewy factor of 0.075.
The computational domain is discretised using \texttt{Carpet} adaptive mesh refinement \citep{Schnetter2004CQGra, Reisswig2013PhRvD} with seven levels of nested Cartesian grids and a factor-of-two refinement between successive grid levels. The finest refinement level covers both \acp{NS} during the inspiral and the central remnant after merger. The numerical resolution is set by the grid spacing of the innermost level, which is $\Delta x \simeq 246~\mathrm{m}$ for all models. To assess the impact of the grid spacing on the nucleosynthesis outcome, we also evolve the SFHo model with $M_1 = M_2 = 1.298\,M_\odot$ at a higher resolution of $\Delta x \simeq 185~\mathrm{m}$ (model \texttt{SFHo$\_$M12980-12980$\_$M1$\_$HR} in \reftab{tab:bns_sims}).
The computational domain extends to $\pm1512~\mathrm{km}$ in each spatial direction. To guarantee numerical stability in low-density regions, we set an artificial atmosphere throughout the computational domain with floor density $\rho = 6\times10^3~{\rm g~cm^{-3}}$.

Nuclear matter is modelled using finite-temperature, composition-dependent \acp{EOS}, specifically the HS(DD2) \ac{EOS} \citep[][hereafter DD2]{Typel2010PRC, Hempel2010NPhysA} and the SFHo \ac{EOS} \citep{Steiner2005PhysRep}. The modelling of nuclear interactions in both \acp{EOS} is based on a relativistic mean field theory, and the relevant degrees of freedom are neutrons, protons, nuclei, electrons, positrons, and photons. Both \acp{EOS} are consistent with current experimental constraints on nuclear matter properties \citep{Shlomo2006EPJA, Danielewicz2013NPA, Oertel2017RvMP, Drischler2017PRL} 
and are in agreement with present astrophysical constraints on the \ac{NS} maximum mass $M_{\rm max} > 2.01 M_\odot$ and radii \citep{Antoniadis2013Science, Lattimer2014APJ, Oertel2017RvMP, Riley2019ApJ, Miller2019ApJ, Miller2021ApJL, Raaijmakers2021APJL}. SFHo produces more compact \acp{NS} compared to DD2: for cold, non-rotating \acp{NS}, SFHo yields a maximum mass of $M_{\rm max}=2.06~M_\odot$ and corresponding radius $R_{\rm max}=10.3$ km, both smaller than the DD2 values of $M_{\rm max}=2.42~M_\odot$ and $R_{\rm max}=11.9$ km. These two \acp{EOS} allow us to span a relatively broad range of \ac{NS} compactness within present uncertainties.

Neutrino transport is handled with a truncated, two-moment energy-integrated radiation transport scheme through the module THC$\_$M1  \citep{Radice2022MNRAS} implemented in \texttt{WhiskyTHC}. The scheme evolves the first
two energy-integrated moments of the neutrino distribution function (energy density and momentum density flux) retaining all nonlinear neutrino-matter coupling terms. The Minerbo closure is applied, ensuring consistency between neutrino diffusive and free-streaming regimes. We take into account three independent neutrino species: electron neutrinos $\nu_e$, electron antineutrinos $\bar{\nu}_e$, and a collective species $\nu_x$, representing muonic and tauonic neutrinos and antineutrinos. Weak interaction rates include charged-current, thermal, and scattering processes as described in previous works \citep[e.g.][]{Galeazzi2013PhRvD,Perego2019EPJA}.

\begin{table*}
  \caption[]{List of the BNS merger simulations considered in this work.}
  \label{tab:bns_sims}
  \centering
  \begin{tabular}{lccccccl}
  \hline \\ [-1.5ex]
  EOS & $M_{1}$ & $M_{2}$ & $q$ & $M_{\rm tot}$ & $t_{\rm mrg}$ & $t_{\rm end}$ & Simulation name\\
   & $[M_\odot]$ & $[M_\odot]$ & & $[M_\odot]$ & [ms] & [ms] & \\
  \hline \\ [-1.5ex]
  DD2 & 1.0899& 1.5570 & 0.70 & 2.6469 &9.17 & 70.99 & \texttt{DD2$\_$M10899-15570$\_$M1} \\
  DD2 & 1.1968& 1.4080 & 0.85 & 2.6048 &10.08 & 65.33 & \texttt{DD2$\_$M11968-14080$\_$M1} \\
  DD2 & 1.2075& 1.7250 & 0.70 & 2.9325 &7.42 & 73.83 & \texttt{DD2$\_$M12075-17250$\_$M1} \\
  DD2 & 1.2980& 1.2980 & 1.00 & 2.5960 &10.18 & 73.14 & \texttt{DD2$\_$M12980-12980$\_$M1} \\
  DD2 & 1.3260& 1.5600 & 0.85 & 2.8860 &7.83 & 74.65 & \texttt{DD2$\_$M13260-15600$\_$M1} \\
  DD2 & 1.4350& 1.4350 & 1.00 & 2.8700 &7.88 & 88.90 & \texttt{DD2$\_$M14350-14350$\_$M1} \\
  SFHo & 1.2980& 1.2980 & 1.00 & 2.5960 &11.59 & 86.00 & \texttt{SFHo$\_$M12980-12980$\_$M1} \\
  SFHo & 1.2980& 1.2980 & 1.00 & 2.5960 &12.38 & 63.55 & \texttt{SFHo$\_$M12980-12980$\_$M1$\_$HR} \\
  \hline \\ [-1.5ex]
  \end{tabular}
  \tablefoot{The equation of state is listed under EOS. $M_1$ and $M_2$ denote the gravitational masses of the binary components, $q = M_1/M_2$ is the mass ratio, $M_{\rm tot}$ is the total mass of the binary, $t_{\rm mrg}$ is the merger time, and $t_{\rm end}$ is the final simulation time.}
\end{table*}

\begin{figure*}[htb!]
    \centering
    \includegraphics[scale = 0.6]{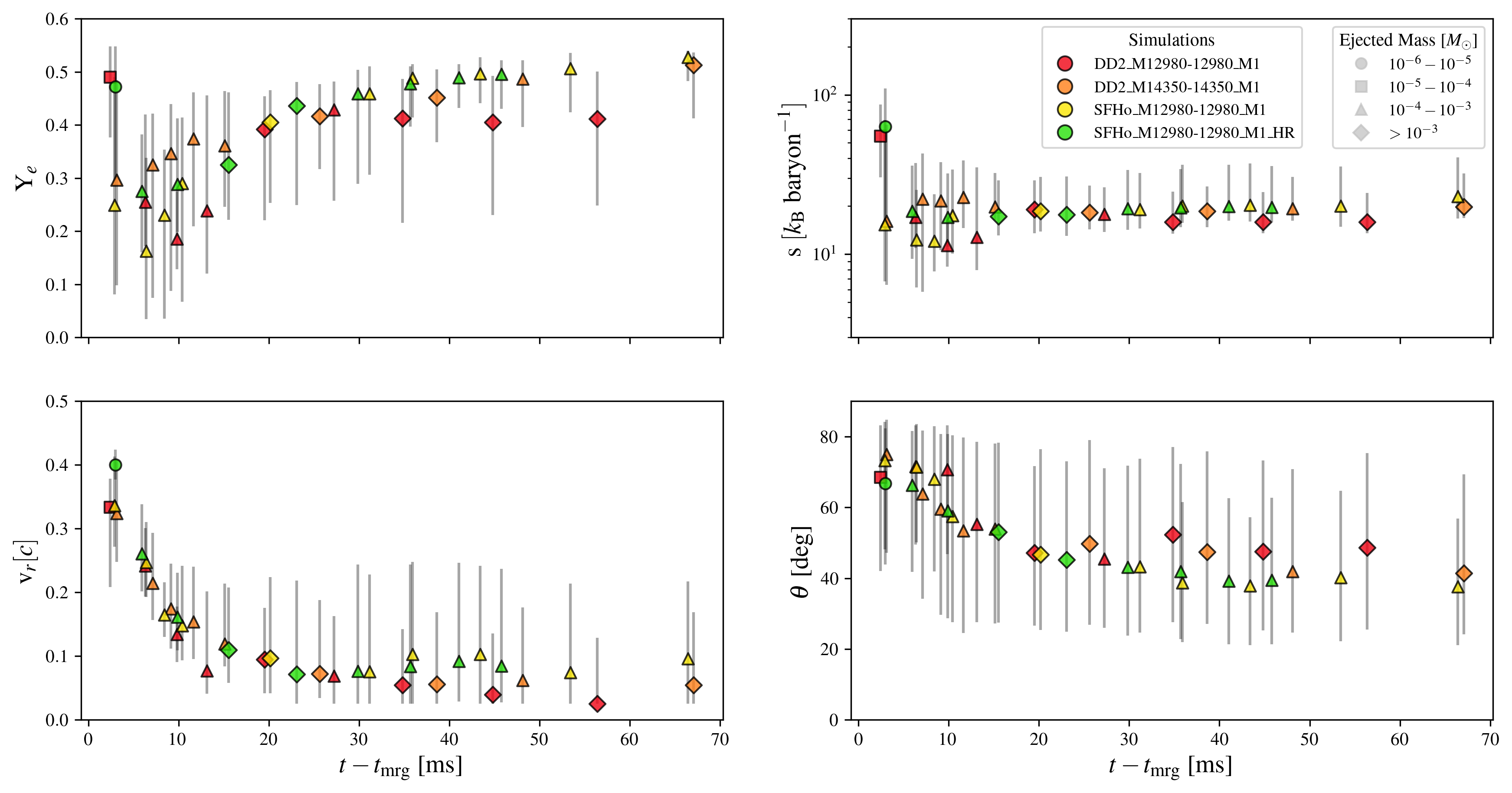}
    \caption{Evolution of the properties of the ejected matter for the $q=1$ simulations as a function of post-merger time ($t - t_{\rm mrg}$), recorded at the detector surface $r_{\rm det} \simeq 443$ km. Each marker represents the median value of the electron fraction $Y_e$ (top left), specific entropy $s$ (top right), radial velocity $v_r$ (bottom left), and polar angle $\theta$ (bottom right) calculated within individual time windows $\Delta t$. Markers are positioned at the centre of their respective time windows. Marker colours indicate the BNS merger simulation while marker shapes indicate the amount of mass ejected during each time interval. Vertical error bars indicate the 16th and 84th percentiles of the distribution within each window. Colours refer to different simulations while markers denote the range of mass ejected during each interval.}
    \label{fig:comparison_all_sims_1}
\end{figure*}

\begin{figure*}[htb!]
    \centering
    \includegraphics[scale = 0.6]{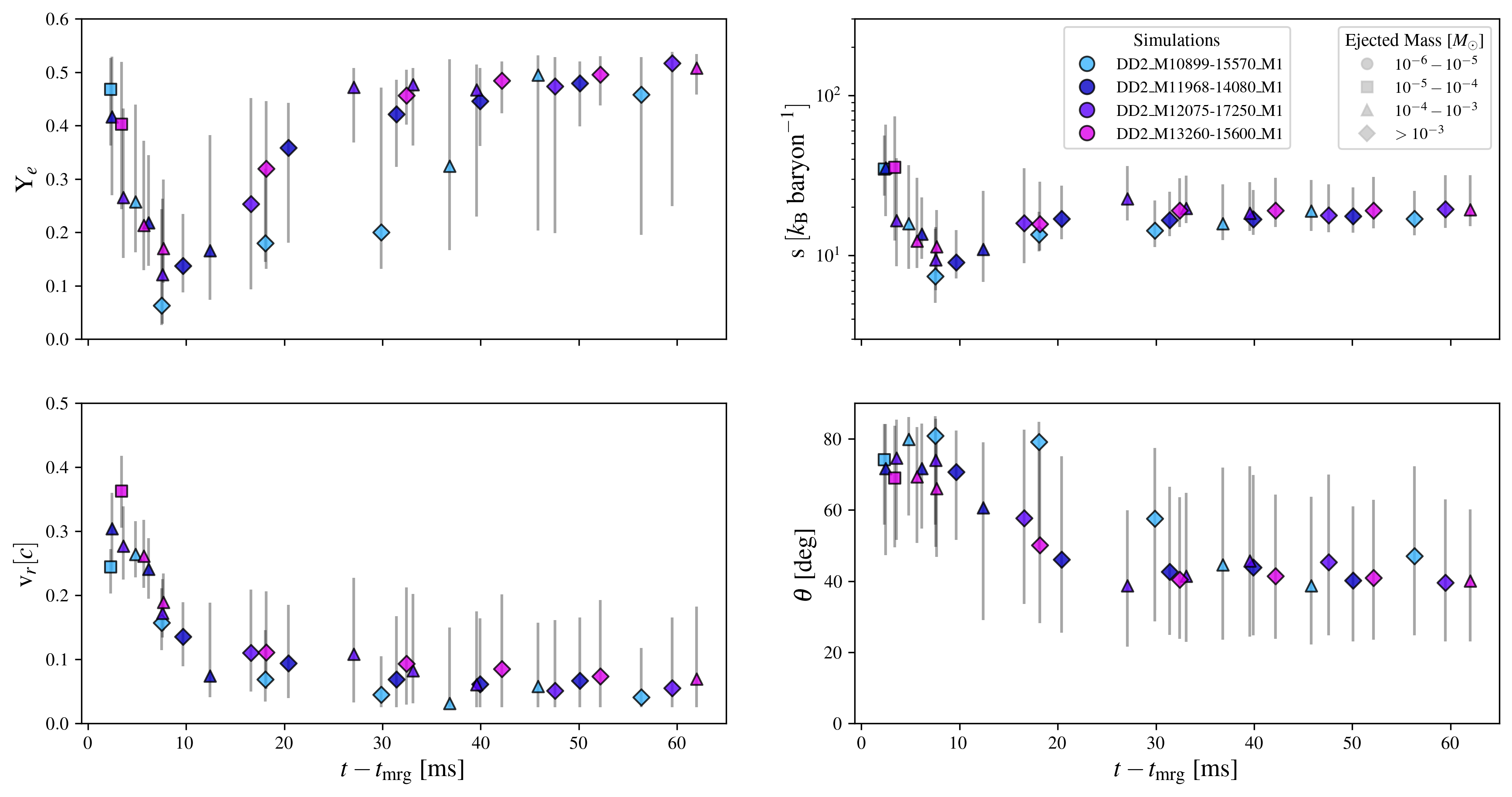}
    \caption{Same as in \reffig{fig:comparison_all_sims_1}, but for BNS simulations with asymmetric mass ratio ($q < 1$).}
    \label{fig:comparison_all_sims_2}
\end{figure*}

\subsection{Ejected matter}\label{subsec:ejected_matter}

To compute the properties of the ejected matter, we combine two complementary approaches. The first involves computing the flux of ejected matter through a spherical surface located at a radius of $r_{\rm det} = 300 \ G M_\odot / c^2 \simeq 443~\rm{km}$ from the center of mass of the system. We define ejected matter as matter that satisfies the Bernoulli criterion $|h~u_t| > 1$, where $u_t$ is the time component of the four velocity and $h$ the specific enthalpy, and has a positive radial velocity $v_{r} > 0$ at the detector surface $r = r_{\rm det}$. We compute the total ejected mass, $m_{\rm ej}$, by integrating the flux from the merger time, $t_{\rm mrg}$, to the end of the simulation, $t_{\rm end}$.
To distinguish between dynamical and wind ejecta, we analyse the mass ejection rate, $\dot{m}(t)$, at $r = r_{\rm det}$. We define the dynamical ejecta mass, $m_{\rm dyn}$, as the mass accumulated up to the first minimum following the first maximum in $\dot{m}(t)$. The wind ejecta mass is defined as the remainder of the ejected matter. We notice that with this definition, the dynamical ejecta mass computed via the Bernoulli criterion practically coincides with that derived from the geodesic criterion (i.e. $|u_t| > 1$). However, as expected, the subsequent wind ejecta mass is significantly larger when using the Bernoulli criterion compared to the geodesic one \citep[e.g.][]{Foucart2021PhRvD}.

Since in all our models the ejecta mass has not saturated by the end of the simulation, the computed wind mass, and consequently the total ejected mass, should be considered as lower limits.
However, since the remnant \ac{NS} has not collapsed yet to a \ac{BH}, we extrapolate the properties of the ejecta over timescales longer than the simulated ones, but comparable to the disc lifetime, by assuming that the disc wind has reached a stationary state.
The additional wind ejecta mass is estimated as a varying fraction of the disc mass, spanning present uncertainties in the long-term post-merger remnant evolution:
\begin{equation}\label{eq:mwind}
    m_{\text{wind}}^{n} = m_{\rm wind,~sim} + \dfrac{n}{10} \left( m_{\rm disc} - m_{\rm disc}^{\rm bern} \right) \quad \text{with} \quad n \in \{0, 1, 2, 3, 4\},
\end{equation}
where $m_{\rm wind, ~sim}$ is the wind ejecta mass computed up to the end of the simulation, $m_{\rm disc}$ is the disc mass at time $t = t_{\rm end}$, $m_{\rm disc}^{\rm bern}$ is the portion of the disc mass satisfying the Bernoulli criterion at $t = t_{\rm end}$.
In each simulation, $m_{\rm disc}$ is calculated by integrating the conserved density over the computational domain in the region where the rest mass density $\rho < 10^{13} \rm{g}~\rm{cm}^{-3}$. \reftab{tab:ejecta} lists, for each simulation, the dynamically ejected mass $m_{\rm dyn}$ and the final disc mass. It also reports the wind ejecta mass for the different wind prescriptions, with minimum $m_{\rm wind}^{n=0}$ and maximum $m_{\rm wind}^{n=4}$, and the corresponding total ejected mass 
\begin{equation}
 m_{\rm ej}^{n} = m_{\rm dyn} + m_{\rm wind}^{n} , 
\end{equation}
with minimum $m_{\rm ej}^{n=0}$ and maximum $m_{\rm ej}^{n=4}$.

The second approach involves tracking passively advected tracer particles. Each simulation initializes $2^{18}$ tracers approximately 2~ms prior the merger, distributed uniformly in space within a radius $r \leq 100 \ M_{\odot}$. Each tracer records thermodynamic and kinetic variables every $\sim 0.06$ ms, and in particular temperature $T$, entropy $s$, rest mass density $\rho$, radius $r$, radial velocity $v_r$, and electron fraction $Y_e$. 
We define tracers associated with the ejecta as all tracers crossing $r_{\rm det}$ that satisfy the Bernoulli criterion and have $v_r > 0$ outside $r_{\rm det}$.

\subsection{Tracer particles selection}\label{subsec:trcrs_part}
To perform the nucleosynthesis calculations, we downsample the full tracer dataset by selecting $3000 - 5000$ representative tracers for each merger model. This sample size is sufficient to accurately reproduce the nucleosynthesis properties of the outflow, provided that the tracers are properly sampled \citep{Ricigliano2024MNRAS2}. To this end, we developed a sampling algorithm designed to preserve the properties of the ejecta. 

First, we divide the outflow flux through $r_{\rm det}$ into eight to ten time windows, $\Delta t$. We set $\Delta t \sim 2 - 5$ ms during the dynamical phase and $\Delta t \sim 5 - 10$ ms during the wind phase. Within each time window, we record the entropy, $s$, and the electron fraction, $Y_e$, of all ejecta crossing the detector surface. These data are then binned in the $Y_e - s$ plane using 28 bins linearly spaced in $Y_e \in [0,0.65]$ and 25 bins logarithmically spaced in $s \in [0.5, 300]~k_{\rm B} {\rm baryon}^{-1}$. Next, we identify all tracers that cross the detector surface within the given time window and satisfy the Bernoulli criterion. Across all simulations, the number of tracers crossing the detector during the wind phase is $\sim 2000-3000$, corresponding to an average tracer mass of $\sim 2\times10^{-6}\,M_\odot$. Therefore, we retain all wind-phase tracers without subsampling. During the dynamical phase, however, the number of crossing tracers is up to two orders of magnitude larger, so we subsample them as follows.
Given the total mass ejected during the dynamical phase, we first estimate the target number of dynamical tracers to retain, $N_{\rm tr}^{\rm dyn}$, by requiring an average tracer mass of $\sim 5\times10^{-7}\,M_\odot$. Each dynamical-phase tracer is then assigned, within its crossing-time window, to a $(Y_e, s)$ bin based on its values at the crossing time. For each dynamical-phase time window, we define the number of tracers to select as
\begin{equation}
    N_{\rm sel} = \max\left(N_{\rm occ},\ \mathcal{N}\right), \quad {\rm where} \quad \mathcal{N} = N_{\rm tr}^{\rm dyn}\,\frac{m_{\Delta t}}{m_{\rm dyn}},
\end{equation}
$N_{\rm occ}$ is the number of bins occupied by at least one tracer in that window, and $m_{\Delta t}$ is the mass ejected within the window. We first select one tracer at random from each occupied bin, guaranteeing full $(Y_e,s)$ coverage. If $N_{\rm occ} < \mathcal{N}$, we then randomly select additional $\mathcal{N}-N_{\rm occ}$ tracers, weighting the selection probability by the ratio $m_{j}/\sum_{j = 1}^{N_{\rm occ}} m_{j}$, where $m_{j}$ is the bin mass. This procedure preserves the $Y_e$-$s$ distribution of the outflow. We verified that this selection also provides a consistent sampling of the ejecta distributions in polar angle, $\theta$, and radial velocity, $v_r$.
We find that the tracer temperature at the detector surface is always $T < 8.5$ GK. The number of selected tracers, $N_{\rm tr}$, for each simulation is reported in \reftab{tab:ejecta}.

To compute the nucleosynthesis yields from each \ac{BNS} merger model, we assign a representative mass to each selected tracer. For each time window, we bin the ejected mass in the three-dimensional parameter space $(Y_e,~s,~\theta)$ at $r=r_{\rm det}$, where the polar angle $\theta$ is defined such that $\theta = 0$ deg at the pole and $\theta = 90$ deg at the equator. We adopt a coarser binning than in the subsampling step above, to ensure adequate tracer statistics per three-dimensional cell: 14 bins linearly spaced in $Y_e \in [0,0.65]$, 10 bins logarithmically spaced in $s \in [0.5, 300]~k_{\rm B}~{\rm baryon}^{-1}$, and 26 bins linearly spaced in $\theta \in [0, \ 90]$ deg. The selected tracers are assigned to the corresponding $(Y_e, s, \theta)$ bins according to their properties at $r = r_{\rm det}$. If a massive cell contains no tracers, we redistribute its mass among the $3\times3\times3 - 1 =26$ neighboring cells in the three-dimensional parameter space until every massive cell is populated by at least one tracer. The mass associated with each populated cell is then divided equally among the tracers it contains. 
We validate this mass-assignment procedure by comparing the ejecta mass distributions as functions of $Y_e,~s,~v_r$, and $\theta$ obtained directly from the outflow flux with those reconstructed from the mass-weighted tracers. The results, shown in \reffig{fig:mass_distr_all_check}, demonstrate a good agreement between the two approaches.

\refeq{eq:mwind} introduces five estimates of the total wind ejecta mass. The first ($n = 0$) uses only the mass directly ejected during the simulation, while the remaining estimates ($n \geq 1$) include an additional contribution from the accretion disc. For $n \geq 1$, the excess mass, $m_{\rm wind} - m_{\rm wind, sim}$, is distributed among the tracers ejected during the final available time window, corresponding to the last $10$  ms of the simulation. 

\begin{table*}[htb!]
  \caption[]{Summary of the outcome of the \ac{BNS} merger simulations considered in this work.}
  \label{tab:ejecta}
  \centering
  \begin{tabular}{lcccccc}
  \hline \\ [-1.5ex]
  Simulation name & $m_{\rm dyn}$ & $\left[m_{\rm wind}^{n = 0},~m_{\rm wind}^{n = 4} \right]$ & $m_{\rm disc}$ & $\left[m_{\rm ej}^{n = 0},~m_{\rm ej}^{n = 4} \right]$& $\left[X_{\rm A > 69}^{n = 0},~X_{\rm A > 69}^{n = 4} \right]$  &$N_{\rm tr}$\\
    &  $[10^{-3} M_\odot]$ & $[10^{-3} M_\odot]$ & $[M_\odot]$ & $[10^{-3} M_\odot]$ & & \\
  \hline \\ [-1.5ex]
  \texttt{DD2$\_$M10899-15570$\_$M1} & 1.772 & $[16.86, 126.7]$ &  0.307 & $[18.63,~128.4]$ & $[0.833, 0.437]$ & 4530\\
  \texttt{DD2$\_$M11968-14080$\_$M1} &1.746 & $[7.086,~110.1]$ &0.283 & $[8.833,~111.8]$ & $[0.632, 0.224]$ & 3417\\
  \texttt{DD2$\_$M12075-17250$\_$M1} & 1.160 & $[10.82,~116.4]$ & 0.284 & $[11.98,~117.6]$ & $[0.616, 0.267]$ & 4032\\
  \texttt{DD2$\_$M12980-12980$\_$M1} &1.049 & $[6.147,~89.38]$&0.232 & $[7.197,~90.43]$ & $[0.595, 0.481]$ & 4086\\
  \texttt{DD2$\_$M13260-15600$\_$M1} & 1.136 & $[10.81,~107.5]$ &0.260 & $[11.94,~108.6]$ & $[0.531, 0.099]$ & 3415\\
  \texttt{DD2$\_$M14350-14350$\_$M1} & 0.599 & $[9.609, 77.81]$ &0.182& $[10.21,~78.41]$ & $[0.407, 0.168]$ & 3321\\
  \texttt{SFHo$\_$M12980-12980$\_$M1} &1.208 & $[8.111,~83.05]$ &0.197 & $[9.319,~84.26]$ & $[0.456, 0.103]$ & 2861\\
  \texttt{SFHo$\_$M12980-12980$\_$M1$\_$HR} &1.121 &$[6.026,~74.65]$&0.185& $[7.147,~75.77]$ & $[0.490, 0.118]$ & 3011\\
  \hline \\ [-1.5ex]
  \end{tabular}
  \tablefoot{$m_{\rm dyn}$, $m_{\rm wind}$, $m_{\rm disc}$, and $m_{\rm ej}$ represent the masses of the dynamical ejecta, wind ejecta, disc, and total ejecta. For the wind ejecta and the total ejecta, we provide the minimum and maximum amount of ejected matter, as discussed in \refsec{subsec:ejected_matter}.  $X_{A>69}^{n=0}$ and $X_{A>69}^{n=4}$ are the mass fractions of $r$-process nuclei ($A>69$) in the total ejecta at $t=1$~Gyr, for the minimum ($n=0$) and maximum ($n=4$) ejected mass, respectively. Finally, $N_{\rm tr}$ represents the number of selected tracers.}
\end{table*}

\subsection{Nucleosynthesis calculations}\label{sec:method_nuclear}

\begin{figure*}[]
    \centering
    \includegraphics[scale = 0.45]{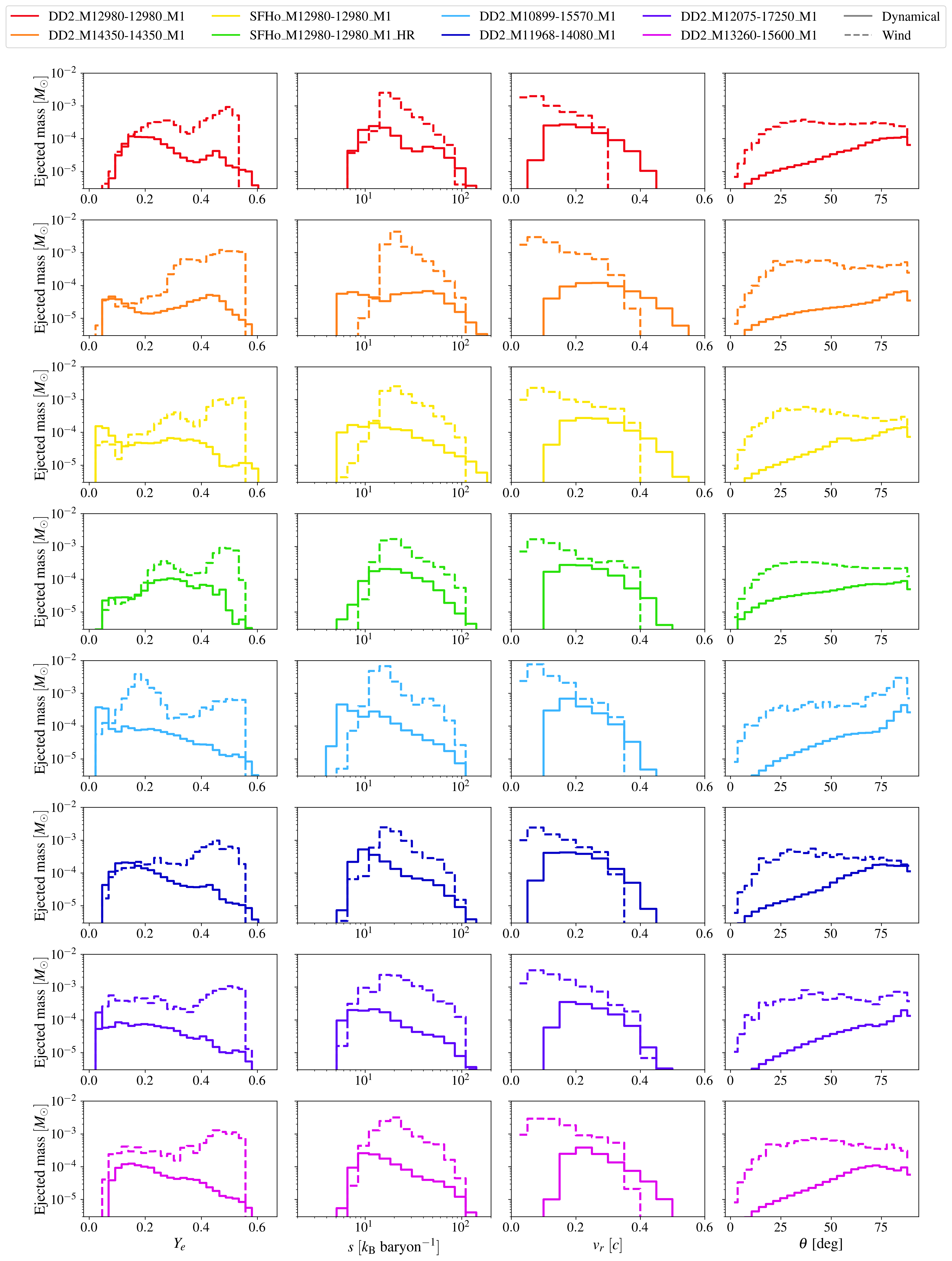}
    \caption{Dynamical (solid lines) and wind (dashed lines) mass distribution in $Y_e$, $s$, $v_r$, and $\theta$ for all simulations reported in \reftab{tab:ejecta}.}
    \label{fig:mass_distr_all}
\end{figure*}

To analyse nucleosynthesis, we post-process the thermodynamic evolution of the selected tracers through the open-source nuclear reaction network \texttt{WinNet}  \citep{Reichert2023ApJS} and evolve abundances up to $t = 1$ Gyr. We provide \texttt{WinNet} with the time-dependent evolution of temperature, density, radius, and electron fraction for each tracer. We set $T = 7$ GK as the threshold temperature below which the full network nucleosynthesis calculation starts, with the initial composition determined by \ac{NSE}. From that point on, the network follows the tracer’s hydrodynamical density $\rho(t)$ and radius $r(t)$ until the end of the trajectory. Afterwards, the density decreases as $\rho\propto t^{-3}$ and the radius grows linearly at the velocity of the last hydrodynamical output interval, i.e. we assume homologous expansion until $t=1$~Gyr. The temperature and the composition are evolved by the network. We do not explicitly include neutrino reactions in the network simulation. We notice that the impact of neutrinos in determining the ejecta electron fraction is already accounted for through M1 neutrino transport in the hydrodynamical simulation. When the network calculations are initialised ($T < 7$ GK), the ejecta have expanded and the neutrino flux is significantly reduced, so that neutrino absorption reactions have a subdominant effect on the final nucleosynthesis\footnote{However, we refer the reader to \cite{Xiong2024PhRvL} for the role of neutrinos in the production of p-nuclei through the $\nu r$-process.}. 

The network includes 8904 nuclei up to $^{339}$Og. We employ the \texttt{REACLIB} database \citep{Cyburt2010ApJS} as our baseline for reaction rates, prioritizing experimental data whenever available. Nuclear properties are adopted from the AME2020 evaluation \citep{Wang2021ChPhC..45c0003W}, complemented with theoretical values derived from the FRDM2012 model \citep{Moller2016ADNDT}. Using the \texttt{TALYS} code \citep{Koning2023EPJA} with recommended parameters from \cite{Rochman2025NuPhA105322951R}, we consistently calculate neutron-capture rates, as well as $\rm{(n, \gamma)}$, $\rm{(n, \alpha)}$, and $\rm{(n, p)}$ channels, based on the same FRDM2012 mass model. Inverse rates are obtained from detailed balance.
Beta decay rates are taken from \cite{Moller2019ADNDT.125....1M}. Additionally, for temperatures larger than $0.01$ GK, we use temperature- and density-dependent theoretical weak interaction rates (including $\beta^-$ and $\beta^+$ decays, as well as positron and electron captures) as described in \cite{Reichert2023ApJS}. Furthermore, $\beta$-delayed fission rates and neutron-induced fission rates are taken from \cite{Panov2005NuPhA.747..633P} and  \cite{Panov2010A&A...513A..61P}, respectively. Spontaneous fission rates are obtained using the semi-empirical formula of \cite{Khuyagbaatar2020NuPhA100221958K}, adopting the fission barriers provided in \cite{Moller2015PhRvC..91b4310M}. Fission fragment distributions are adopted from \cite{Mumpower2020PhRvC.101e4607M} for $\beta$-delayed and neutron-induced fission, and from \cite{Kodama1975NuPhA.239..489K} for spontaneous fission. 

In \reftab{tab:ejecta}, we report the mass fraction of $r$-process nuclei ($A>69$) in the total ejecta at $t=1$~Gyr, for the minimum ($n=0$) and maximum ($n=4$) wind ejecta mass, denoted $X_{A>69}^{n=0}$ and $X_{A>69}^{n=4}$, respectively.

\section{Results}\label{sec:results}

\subsection{Ejected matter}\label{sec:ejected_matter}
The time evolution of the ejecta properties reveals several common features across all simulations, particularly in the transition from dynamical to wind ejecta.
This evolution is summarised in \reffig{fig:comparison_all_sims_1} and \reffig{fig:comparison_all_sims_2} for symmetric ($q=1$) and asymmetric ($q < 1$) \ac{BNS} models, respectively. Both figures show the median values and the 16th to 84th percentile distributions of the electron fraction $Y_e$, entropy $s$, radial velocity $v_r$, and polar angle $\theta$ as functions of the post-merger time. 

In all models, the dynamical ejecta are initially ($t-t_{\rm mrg} \lesssim 3~{\rm ms}$) characterised by a small amount of mass ($\delta m_{\rm ej} \lesssim 10^{-5} M_{\odot}$), expanding at high speed ($v_r \gtrsim 0.3~c$) across the equatorial plane, with a high median electron fraction around $\langle Y_e \rangle \sim 0.4-0.5$, and a relatively high entropy 
($s \gtrsim 50~k_{\rm b}~{\rm baryon}^{-1}$ and $s \gtrsim 30~k_{\rm b}~{\rm baryon}^{-1}$ for $q=1$ and $q <1$, respectively). These ejecta originate from the collision of the two \acp{NS}. 

After that, but still within the first $\sim$10~ms after merger, the median electron fraction and the expansion velocity decrease, accounting for the ejection of slower and more neutron rich matter due to tidal interaction and remnant shocks.
The minima of $\langle Y_e \rangle$ are consistently reached during this phase. Specifically, we find that all simulations reach $\langle Y_e \rangle < 0.25$, with the exception of \texttt{DD2$\_$M14350-14350$\_$M1} and \texttt{SFHo$\_$M12980-12980$\_$M1$\_$HR}. The lowest values ($\langle Y_e \rangle < 0.15$) are observed in the two highly asymmetric cases ($q=0.7$), namely \texttt{DD2$\_$M10899-15570$\_$M1} and \texttt{DD2$\_$M12075-17250$\_$M1}, followed by the $q=0.85$ model \texttt{DD2$\_$M13260-15600$\_$M1}.
This behaviour is consistent with previous analysis, showing that the competition between the expansion timescale and the neutrino irradiation effect eventually sets the ejecta composition. Since dynamical ejecta are mostly emitted across the equator (in particular, for very asymmetric models), where the shielding effect due to the presence of the disc is stronger, most of the ejecta do not reach the equilibrium electron fraction, partially preserving their initial neutron richness.
In the subsequent wind phase, $\langle Y_e \rangle$ steadily increases again, finally stabilising at $\langle Y_e \rangle \sim 0.4-0.5$ around 30 ms after merger.
At the same time, both the median entropy and radial velocity stabilise around $\langle s \rangle \approx 20~k_{\rm B}~{\rm baryon}^{-1}$ and $\langle v_r \rangle \leq 0.1~c$, respectively, during wind ejection.
This entire trend shows the growing impact of the neutrino irradiation, which progressively increases the electron fraction of the outflowing matter.
In particular, neutrino absorption provides enough energy to fluid elements to expand and decrease temperature inside the outer disc layers. The subsequent nuclear recombination supplies the necessary energy to unbind matter, setting the final speed and entropy.
Regarding the angular distribution, most of the ejecta progressively migrate toward more polar directions once the neutrino-driven wind develops \citep[cf.][]{Bernuzzi2025MNRAS}, ultimately stabilising at $\langle \theta \rangle \approx 40~ {\rm deg} - 50~ {\rm deg}$ with a relatively broad dispersion $( \pm 20~ {\rm deg})$. 

When comparing \texttt{SFHo$\_$M12980-12980$\_$M1} and \texttt{DD2$\_$M12980-12980$\_$M1}, in the wind ejecta SFHo yields higher electron fraction values and a significantly smaller spread compared to DD2. Notably, while the baseline and high-resolution (\texttt{HR}) runs of the SFHo model show distinct electron fraction behaviours during the dynamical phase, they converge to nearly identical values during the wind phase.
Overall, the dependence of the electron fraction on the binary mass ratio and \ac{EOS}, together with its systematic increase during the wind phase due to neutrino irradiation, is consistent with previous \ac{BNS} merger simulations employing detailed neutrino transport \citep{Radice2023ApJ,Radice2024JPhCS,Bernuzzi2025MNRAS}. Also the variation due to numerical resolution is consistent with previous studies 
\cite[see e.g.][]{Zappa2023MNRAS}.

In \reffig{fig:mass_distr_all}, we show the mass distribution of $Y_e$, $s$, $v_r$, and $\theta$ for the dynamical (solid lines) and wind (dashed lines) ejecta. The largest dynamical ejecta mass is found in the asymmetric ($q = 0.7$), relatively low-mass model \texttt{DD2$\_$M10899-15570$\_$M1}, while the smallest is produced by the equal and high-mass model \texttt{DD2$\_$M14350-14350$\_$M1}. For most simulations, the electron fraction of the dynamical ejecta peaks at $Y_e < 0.2$. However, both the SFHo models peak at higher values around $Y_e \approx 0.3$. Since SFHo is a softer equation of state, it leads to stronger shocks and higher remnant temperatures, enhancing positron captures on neutrons and increasing neutrino fluxes, therefore shifting the $Y_e$ distribution toward higher values. The equal-mass model employing the DD2 EOS with $M_1 = M_2 = 1.4350~M_\odot$ exhibits a bimodal $Y_e$ distribution with peaks at $0.1$ and $0.4$. The first peak originates from the tidal interaction between the two merging stars, enhanced by the EOS stiffness. The second one, coming from the NS contact interface, is amplified by the larger \ac{NS} masses and by the symmetric merger.
This bimodality is mirrored in its entropy distribution, which shows peaks at $s \approx 6~k_{\rm B}~{\rm baryon}^{-1}$ and $s \approx 50~k_{\rm B}~{\rm baryon}^{-1}$, while all other models peak near $s \approx 10~k_{\rm B}~{\rm baryon}^{-1}$. The velocity distribution of dynamical ejecta typically peaks at $v_r \approx 0.2~c$, though \texttt{SFHo$\_$M12980-12980$\_$M1} and \texttt{DD2$\_$M14350-14350$\_$M1} exhibit significant tails reaching up to $0.55~c$. While most matter is emitted near the equatorial plane, the high resolution SFHo displays a broader emission angle, extending toward the poles down to $\theta \approx 25$ deg. 

During the wind phase, the electron fraction distribution generally peaks between $Y_e \approx 0.4$ -- $0.5$ for all simulations. The two asymmetric DD2 models ($q=0.7$) are notable exceptions, showing a double-peaked distribution ($Y_e \approx 0.2$ and $0.5$). Concerning the entropy in the wind, all simulations show a peak near $s \approx 20~k_{\rm B}~{\rm baryon}^{-1}$. The velocity distribution peaks at $v_r \approx 0.1~c$, with the wind phase generally being slower than the dynamical one. Finally, the wind shows peaks at $\theta \approx 30$ deg – $40$ deg for most models, driven by neutrino irradiated winds that shift the emission toward the poles. However, the $q=0.7$ models retain significant equatorial emission, due to their tidally dominated merger dynamics. Specifically, the DD2 simulation with $M_1 = 1.0899~M_\odot$ and $M_2 = 1.5570~M_\odot$ peaks exclusively at the equator, while DD2 with $M_1 = 1.2075~M_\odot$ and $M_2 = 1.7250~M_\odot$, shows a double peak at the equator and around $30$ deg $- \ 40$ deg.

\begin{figure*}[]
    \centering
    \includegraphics[scale = 0.67]{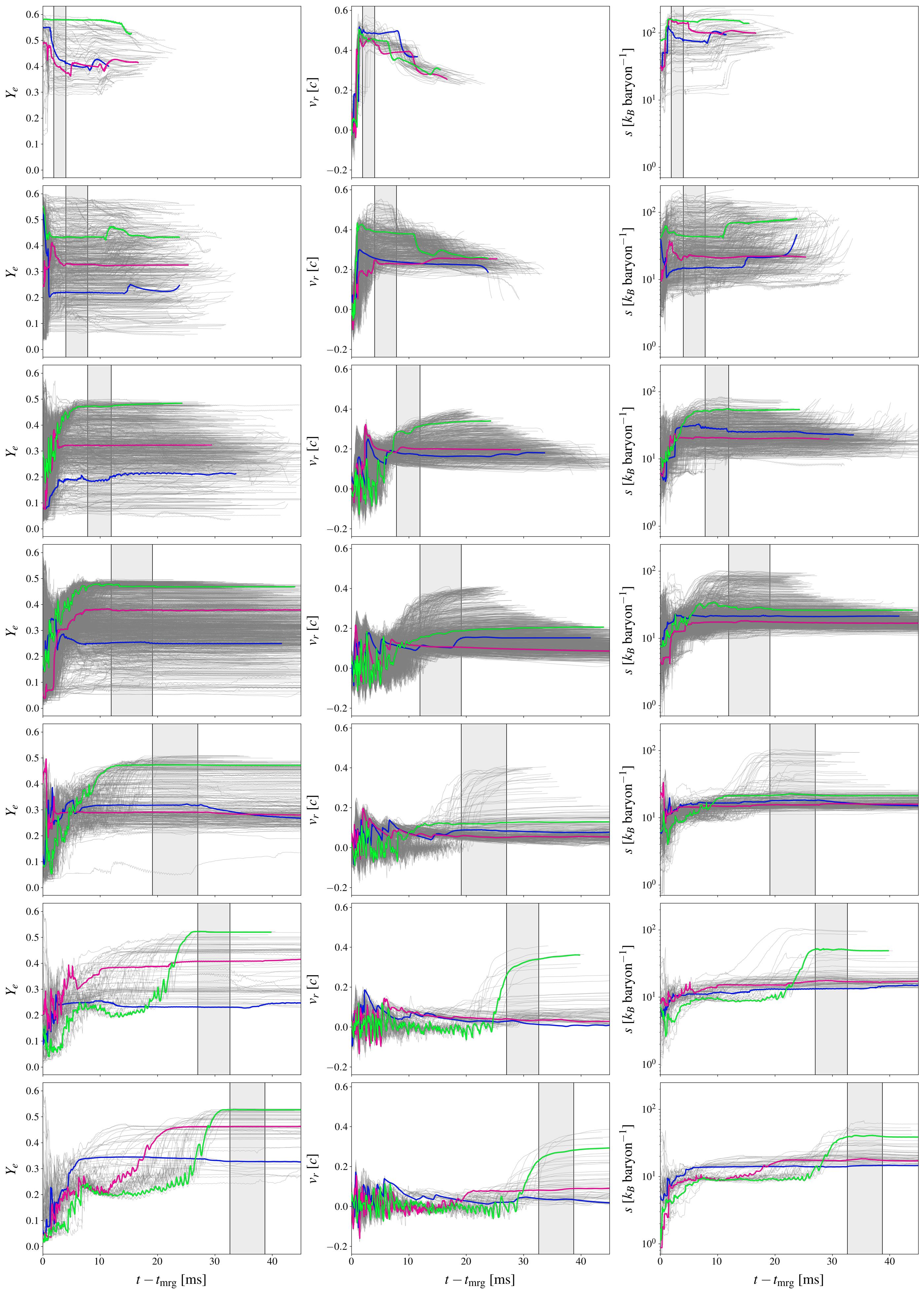}
    \caption{Ejected tracer trajectories (grey) for the electron fraction (left), radial velocity (center), and specific entropy (right) as functions of $t-t_{\rm mrg}$, for the simulation \texttt{SFHo$\_$M12980-12980$\_$M1$\_$HR}. Each row corresponds to a time window $\Delta t$, marked by the shaded grey region bounded by solid black lines; within each row, we plot the trajectories of all tracers crossing the detector surface $r_{\rm det}$ within that window. The blue, magenta, and green lines highlight three representative tracers. The first three rows represent tracers crossing $r_{\rm det}$ during the dynamical phase, while the remaining rows refer to the wind phase up to $\sim 35$ ms after merger.}
    \label{fig:tracers}
\end{figure*}

\subsection{Tracers evolution}

In \reffig{fig:tracers}, we show the time evolution of the electron fraction (left), radial velocity (center), and specific entropy (right) for the tracer particles extracted from the simulation \texttt{SFHo\_M12980-12980\_M1\_HR}. In each panel, we include the trajectories that crossed the detector surface, $r_{\rm det}$, within the time window indicated by the light grey region. Thus, the first three rows represent tracers ejected during the dynamical phase, while the last three rows refer to the early wind phase, up to $\sim 35$ ms after merger. For each time window, three representative tracer trajectories are superimposed in green, blue, and magenta.  Tracer trajectories crossing $r_{\rm det}$ between $\sim 35$ ms and $\sim 51$ ms after merger (end of the simulation) behave as those represented in the last row.

Tracers ejected immediately after merger (first row) originate from the shock interface and exhibit the characteristic behaviour of shocked matter: the radial velocity shows jumps due to shocks and peaks at $\sim 0.5~c$, while the entropy rises above $\sim 50~k_{B}~{\rm baryon}^{-1}$. 
The remaining dynamical ejecta (second and third rows) are characterised by tracers spanning a broad range of electron fractions, $Y_e \sim 0.1-0.5$. These tracers show a rapid velocity increase immediately after merger, followed by an approximately constant radial velocity, and a more gradual rise in entropy. The representative trajectories in these rows highlight different behaviours. Some tracers are ejected ballistically (e.g., blue in the second row), maintaining nearly constant $Y_e$, $s$, and $v_r$ after the initial acceleration. Other tracers show signatures of fluid compression (e.g., the magenta and blue tracers in the third row), leading to increases in both $Y_e$ and $s$. A further class of tracers, characterised by multiple oscillations in $v_r$, is ejected after spending time in the forming remnant disc (e.g., the green tracer in the third row). These tracers exhibit increasing $Y_e$ and $s$ due to neutrino irradiation. 
The impact of neutrino irradiation becomes increasingly significant with the onset of the wind phase, leading to a systematic rise in both $Y_e$ and $s$ as tracers spend time in the disc. Radial velocity oscillations around zero reflect the back-and-forth motion of tracers within the disc prior to their ejection (see, e.g., the green and magenta tracers in the fifth and sixth rows). This behaviour becomes progressively more common at later times.

Tracer trajectories in all other simulations follow similar trends; however, during the dynamical phase, simulations with highly asymmetric mass ratios exhibit a $Y_e$ distribution that is more strongly weighted toward lower values (see also the discussion in \refsec{sec:ejected_matter}). 

\begin{figure}[]
    \centering
    \includegraphics[scale = 0.33]{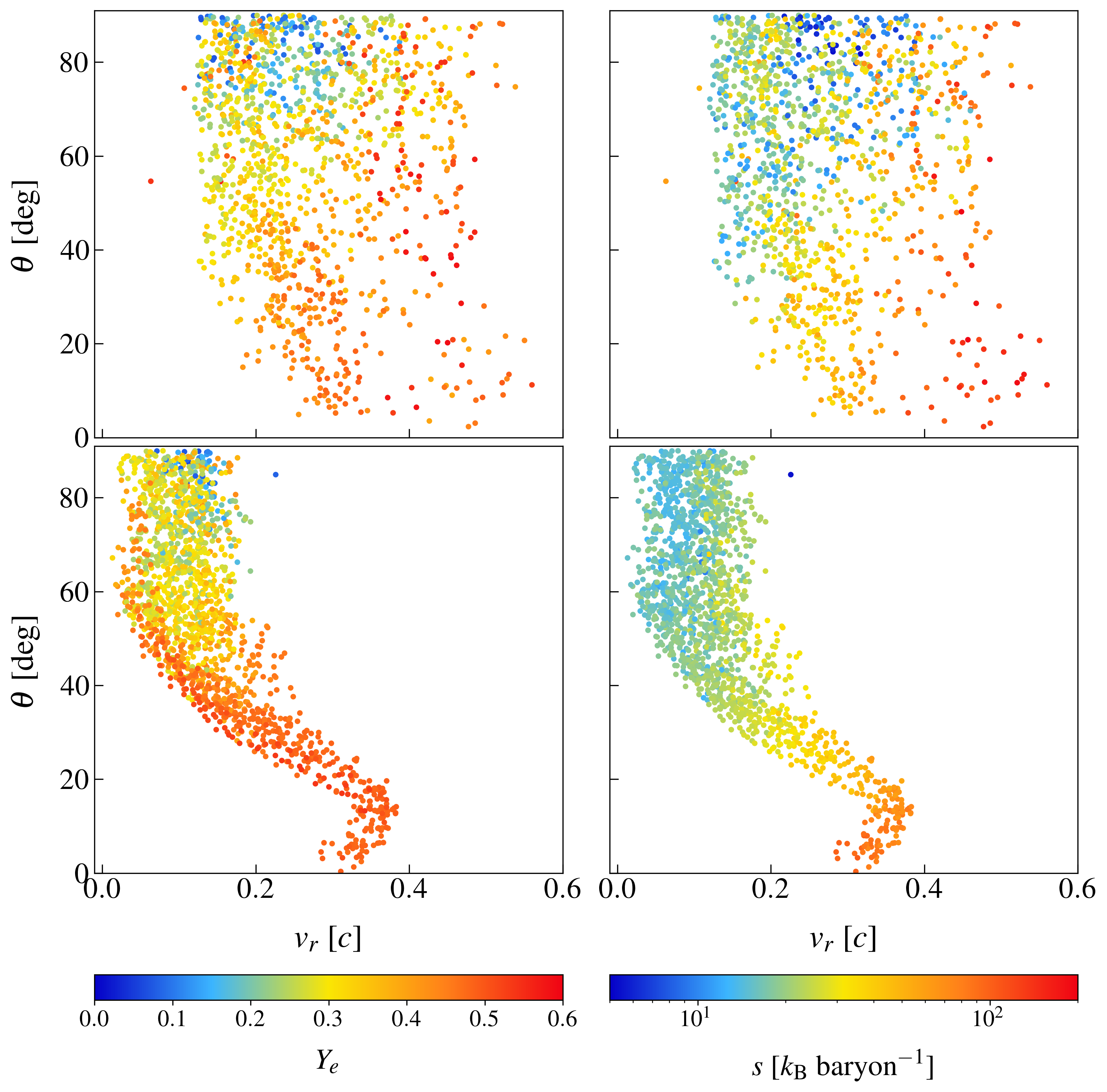}
    \caption{Tracers as function of the radial velocity, $v_r$, and the polar angle, $\theta$, extracted at time $t_{7}$, such that the temperature $T(t) < 7~{\rm GK}$ for all $t > t_{7}$, considering the simulation \texttt{SFHo$\_$M12980-12980$\_$M1$\_$HR}. Upper (lower) row represents tracers crossing the detector surface during the dynamical (wind) phase. Colour bars refer to the electron fraction (left) and the entropy (right).}
    \label{fig:tracers_7GK}
\end{figure}

In \reffig{fig:tracers_7GK}, we present the tracer distribution in the $v_r - \theta$ plane, extracted at time $t_{7}$, defined such that the temperature $T(t) < 7~{\rm GK}$ for all $t > t_{7}$, considering the simulation \texttt{SFHo$\_$M12980-12980$\_$M1$\_$HR}. The upper (lower) row corresponds to tracers that crossed the detector surface during the dynamical (wind) phase. The colour of each tracer represents its electron fraction (left panels) and entropy (right panels). A clear trend emerges in the tracer distribution: tracers with higher electron fraction and entropy are preferentially emitted near the polar regions and exhibit larger radial velocities \citep[e.g.][]{Radice2024JPhCS, Bernuzzi2025MNRAS}. As the electron fraction and entropy decrease, the tracers shift toward the equatorial regions and their radial velocities progressively decrease. This trend is particularly evident for tracers emitted during the wind phase, whereas tracers associated with the dynamical ejecta show a much broader spread\footnote{For example, the highly asymmetric simulation \texttt{DD2$\_$M10899-15570$\_$M1} exhibits a few tracers ejected in the polar region during the dynamical phase with $Y_e < 0.2$ and $s > 100~k_{\rm B}~{\rm baryon}^{-1}$.}.
We also note the presence of dynamical ejecta tracers in the high-velocity tail, reaching $v_r \gtrsim 0.5~c$, consistently with the findings of  \citealt{Radice2018ApJ, Nedora2021ApJ, Bernuzzi2025MNRAS, Rosswog2025MNRAS}. Tracers from the other simulations display the same qualitative correlations between $Y_e$, $s$, $v_r$, and $\theta$, although the detailed shape of the distributions varies from simulation to simulation \footnote{All tracer trajectories produced in this study are publicly available at the following link: \url{https://doi.org/10.5281/zenodo.23689687}.}.

\subsection{Nucleosynthesis}

\subsubsection{Yields at 1 day}\label{sec:yields_1d}

The analysis of AT2017gfo spectra revealed a robust identification of strontium (Sr) features \citep{Watson2019Nature, Domoto2021ApJ}, with an ongoing discussion on the potential contribution of He \citep{Perego2022ApJ, Sneppen2024A&A, Sneppen2026PhRvD, Chiba2026arXiv}, and plausible identifications of features due to yttrium (Y) \citep{Sneppen2023A&A}, tellurium (Te) \citep{Hotokezaka2023MNRAS, Gillanders2024MNRAS}, lanthanum (La) and cerium (Ce) \citep{Domoto2022ApJ}, and gadolinium (Gd) \citep{Gillanders2024MNRAS, Rahmouni2025ApJ}. In addition, the effect of nickel (Ni) \citep{Jacobi2024MNRAS} and Th \citep{Domoto2025ApJ} on kilonova lightcurves and spectra was also recently investigated.

\begin{figure*}[htb!]
    \centering
    \includegraphics[scale = 0.52]{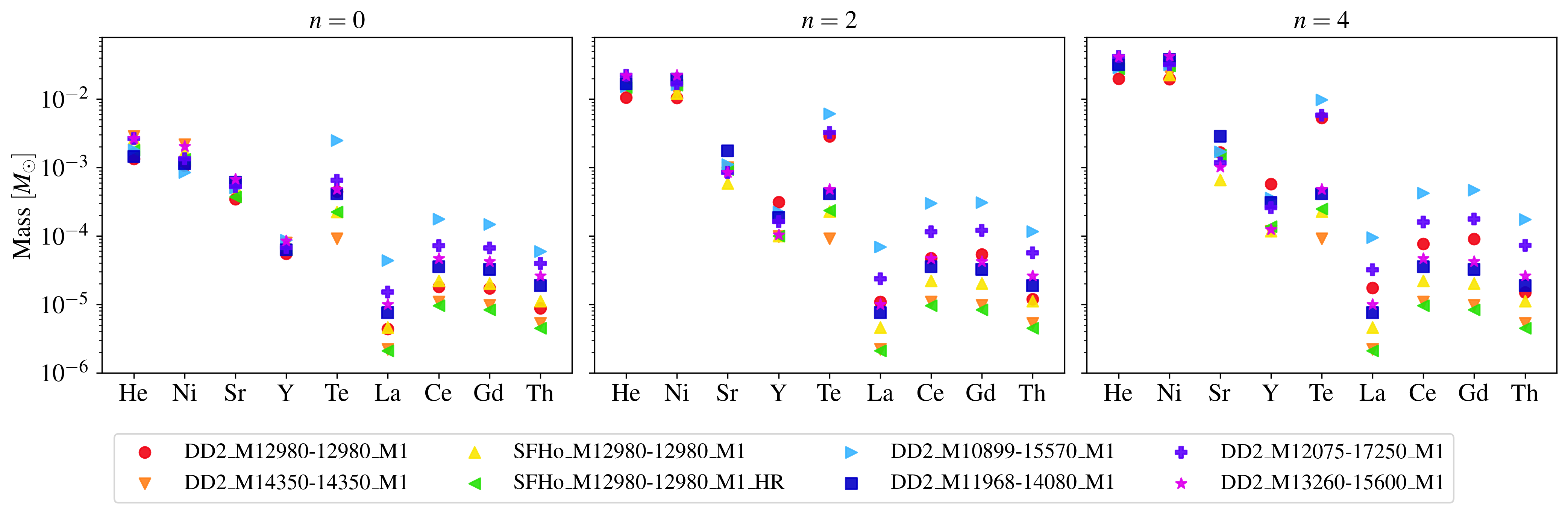}
    \caption{Masses of selected elements synthesised at $t = 1$ day after merger for all simulations listed in \reftab{tab:ejecta}. From left to right, the wind ejecta mass is computed according to \refeq{eq:mwind}, by considering only the wind mass directly ejected during the simulation time ($n = 0$) and by adding $20\%$ or $40\%$ of the accretion disc mass ($n = 2$ and $n = 4$), respectively.}
    \label{fig:ele_1day}
\end{figure*}

In \reffig{fig:ele_1day}, we show the masses of He, Ni, Sr, Y, Te, La, Ce, Gd, and Th present at $t = 1$ day after merger in the ejecta of all simulations in our sample. 
The figure includes three of the five prescriptions for the wind ejecta mass introduced in \refeq{eq:mwind}. We first consider the case $m_{\rm wind} = m_{\rm wind,~sim}$ ($n = 0$, leftmost panel), where only the ejecta directly produced during the simulation are included. In this case, all merger models predict similar yields of the lightest elements, namely He ($1$--$3\times10^{-3}\,M_\odot$), Ni ($0.8$--$2\times10^{-3}\,M_\odot$), Sr ($3$--$7\times10^{-4}\,M_\odot$), and Y ($6$--$8\times10^{-5}\,M_\odot$). In contrast, the production of heavier elements, including Te, La, Ce, Gd, and Th, exhibits a much larger model-to-model variation, spanning more than one order of magnitude. As expected, unequal-mass \acp{BNS} generally produce larger amounts of heavy elements than equal-mass systems. As discussed in \refsec{sec:ejected_matter} and shown in \reffig{fig:mass_distr_all}, these mergers eject a larger fraction of low $Y_e$ and low entropy material due to the reduced contribution of shock-heated ejecta. Such thermodynamic conditions favour the production of heavy $r$-process nuclei. Among the elements shown in \reffig{fig:ele_1day}, He and Ni are the most abundant in all models, with the exception of \texttt{DD2$\_$M10899-15570$\_$M1}, which synthesises more Te ($\sim3\times10^{-3}\,M_\odot$) than either He or Ni.

\begin{figure}[htb!]
    \centering
    \includegraphics[scale = 0.5]{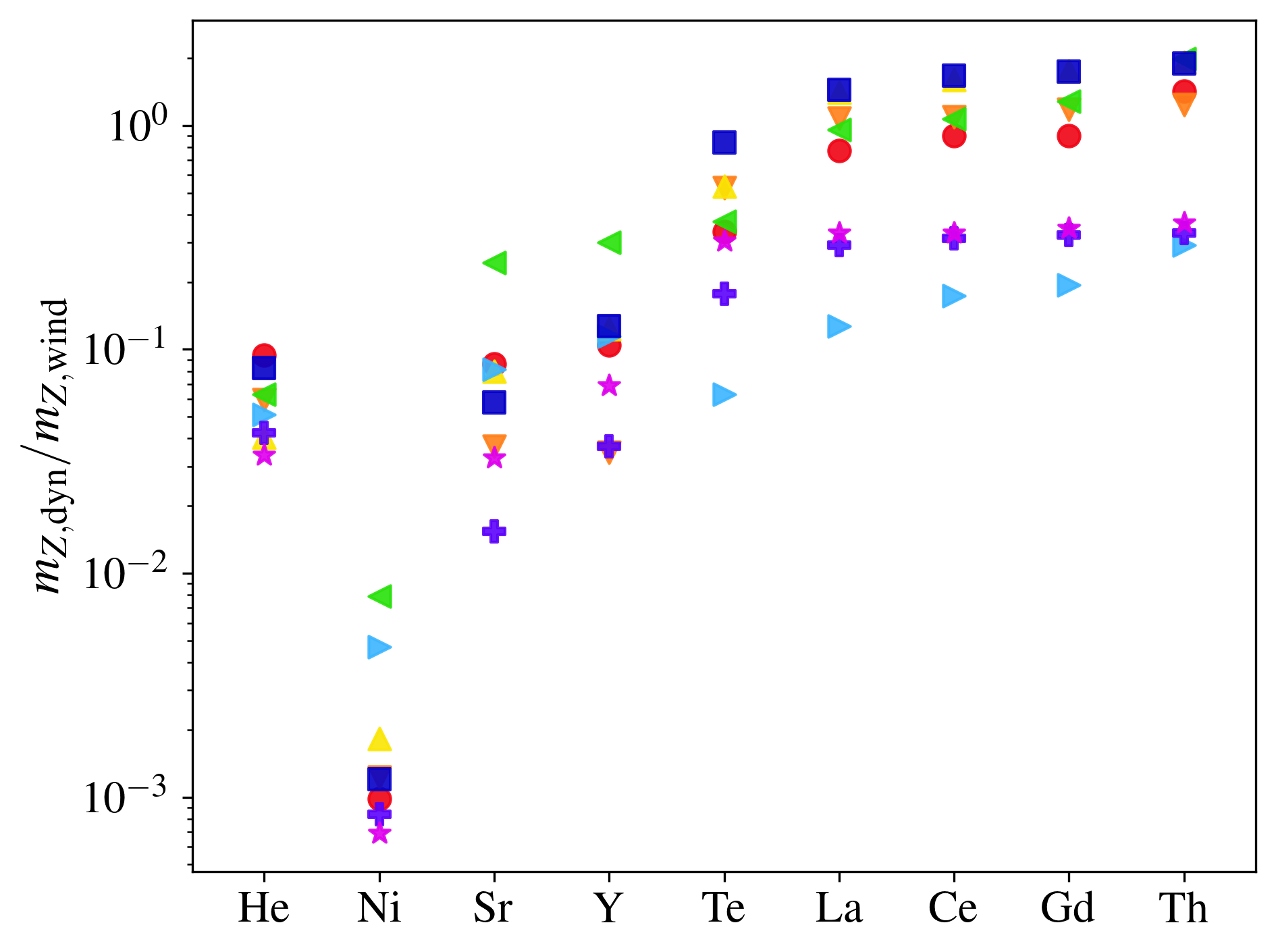}
    \caption{Ratio of the mass of selected elements $Z$ synthesised in the dynamical ejecta, $m_{Z,{\rm dyn}}$, to that synthesised in the wind ejecta, $m_{Z,{\rm wind}}$, at $t = 1$ day after merger, for all simulations listed in \reftab{tab:ejecta}. The wind ejecta mass is computed according to \refeq{eq:mwind} considering only the material directly ejected during the simulation ($n = 0$). The legend is identical to that of \reffig{fig:ele_1day}.} 
    \label{fig:ele_1day_dynwind}
\end{figure}

Increasing the wind mass ($n \geq 1$) enhances the production of all elements. The effect is particularly pronounced for He, Ni, Sr, Y, and Te, while the spread among different merger models also increases. The origin of this behaviour becomes evident from \reffig{fig:ele_1day_dynwind}, where we show the ratio of the elemental yield in the dynamical ejecta to that in the wind ejecta in the case $n=0$.
For every simulation, most of the He, Ni, Sr, Y, and Te are synthesised in the wind ejecta. Consequently, increasing the wind mass primarily boosts the production of these elements. Interestingly, He exhibits only a small model-to-model variation in both the dynamical and wind components, indicating that its production is relatively insensitive to the details of the merger. 
Focusing on nuclei heavier than Te, for most simulations they are predominantly synthesised in the dynamical ejecta and in the early-time wind ejecta, where the material is characterised by lower entropy and lower electron fraction. As a result, increasing the late-time wind mass has only a modest impact on their total yields. However, the two unequal-mass mergers with $q = 0.7$, namely \texttt{DD2$\_$M10899-15570$\_$M1} and \texttt{DD2$\_$M12075-17250$\_$M1}, produce a larger amount of heavy elements in the wind than in the dynamical ejecta, with a significant component of heavy elements coming from the late time wind. The equal-mass model \texttt{DD2$\_$M12980-12980$\_$M1} behaves similarly, synthesizing comparable amounts of heavy nuclei in the two ejecta components, again with a significant late-time wind contribution. This behaviour is consistent with Figs.~\ref{fig:comparison_all_sims_1} and \ref{fig:comparison_all_sims_2}, which show that the late-time wind ejecta of these three models retain a broad $Y_e$ distribution extending to low values. Consequently, increasing the wind mass also significantly enhances the production of heavy nuclei in these simulations, amplifying the model-to-model spread seen in \reffig{fig:ele_1day}. This broad $Y_e$ distribution in the wind ejecta of the more asymmetric simulations is consistent with the spiral-wave wind mechanism, which produces a broad $Y_e$ distribution and is particularly efficient in more asymmetric mergers \citep{Nedora2021ApJ}.
The less strongly asymmetric simulations \texttt{DD2$\_$M11968-14080$\_$M1} and \texttt{DD2$\_$M13260-15600$\_$M1} follow a different pattern: although the amount of heavy nuclei produced in the dynamical ejecta is comparable to or smaller than that produced in the wind ejecta, most of the heavy-nuclei production in the wind occurs at early times. Increasing the late-time wind component therefore does not significantly increase the yield of heavy nuclei in these two models.

\begin{figure*}[htb!]
    \centering
    \includegraphics[scale = 0.47]{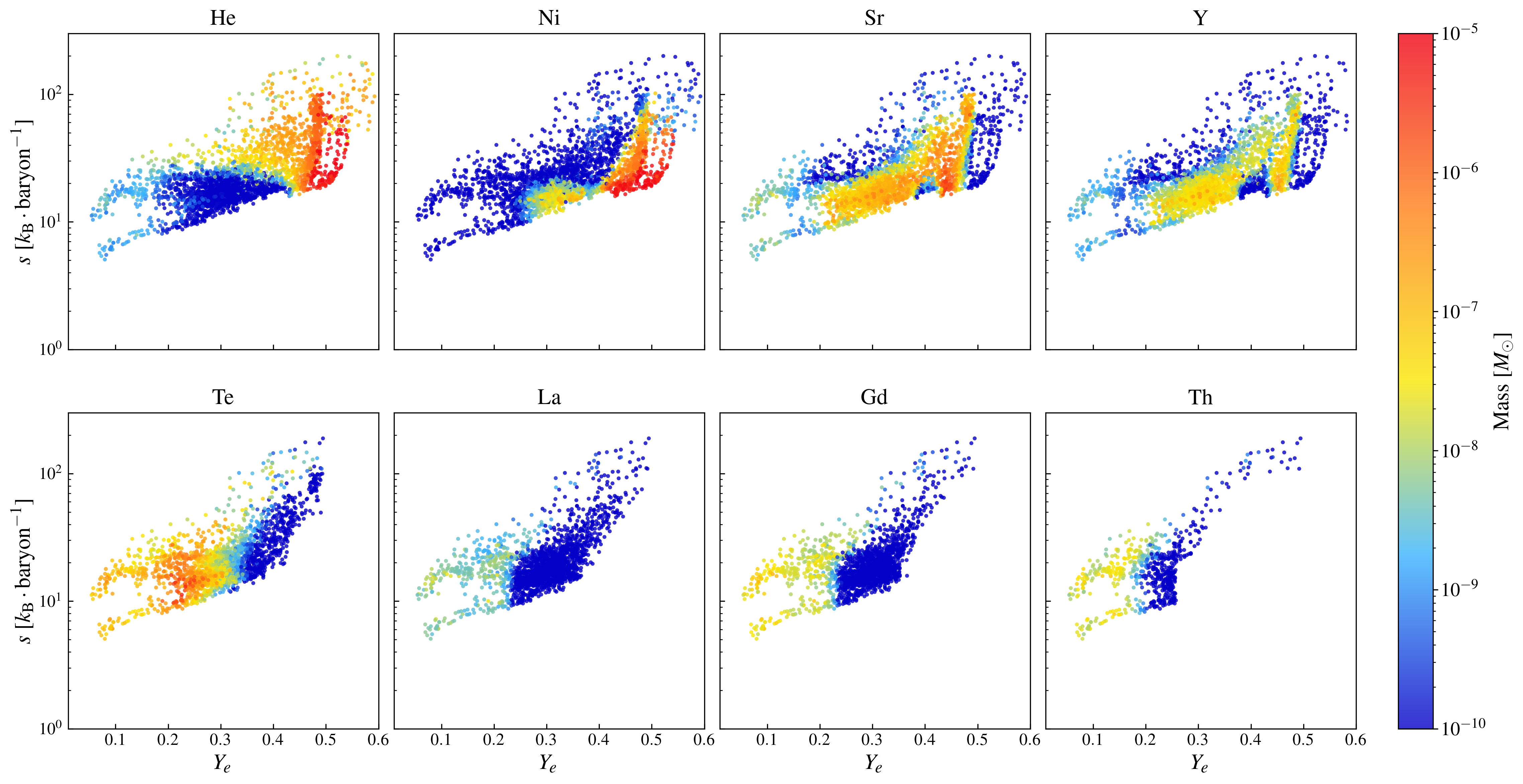}
    \caption{Tracer distribution in the $Y_e$-$s$ plane at time $t_{7}$, defined such that $T(t) < 7~{\rm GK}$ for all $t > t_{7}$, for the simulation \texttt{SFHo$\_$M12980-12980$\_$M1$\_$HR}. The color scale indicates the mass of the element produced by each tracer at $t=1$ day after merger. The element corresponding to each panel is indicated in its title.} 
    \label{fig:origin_ye_s}
\end{figure*}

The thermodynamic conditions under which the different elements are synthesised are illustrated in \reffig{fig:origin_ye_s}. The plots show the tracer distribution in the $Y_e$-$s$ planes at time $t_7$ for the representative simulation \texttt{SFHo$\_$M12980-12980$\_$M1$\_$HR}. They provide a direct visualization of the distinct thermodynamic regimes responsible for the synthesis of light and heavy nuclei.
He is predominantly synthesised in high-$Y_e$ ($Y_e \simeq 0.5$) and high-entropy ($s \gtrsim 30 \ k_{\rm B} \ {\rm baryon}^{-1}$) ejecta, as expected for $\alpha$-rich \ac{NSE} freeze-out. Ni is also mainly produced in the high-$Y_e$ region ($Y_e \gtrsim 0.4$), but a significant fraction also originates from the ejecta with slightly lower electron fractions ($Y_e \simeq 0.3\text{--}0.4$) and lower entropies ($s \sim 10\text{--}20 \ k_{\rm B} \ {\rm baryon}^{-1}$). The production of Sr and Y extends over a much broader region of parameter space, approximately spanning $Y_e \simeq 0.2\text{--}0.5$ and $s \simeq 8\text{--}100 \ k_{\rm B}~{\rm baryon}^{-1}$. In this region, the dominant contribution follows a correlation between the electron fraction and the entropy, with higher electron fractions corresponding to higher entropies. In contrast, the production of heavier nuclei is concentrated in low-$Y_e$, low-entropy ejecta, with $Y_e < 0.3$ and $s < 30 \ k_{\rm B} \ {\rm baryon}^{-1}$. This behaviour is particularly evident for Th, whose production peaks in the region $Y_e < 0.2$ and $s < 30 \ k_{\rm B} \ {\rm baryon}^{-1}$, in agreement with the results of \cite{Eichler2019ApJ}. 

\begin{figure*}[htb!]
    \centering
    \includegraphics[scale = 0.44]{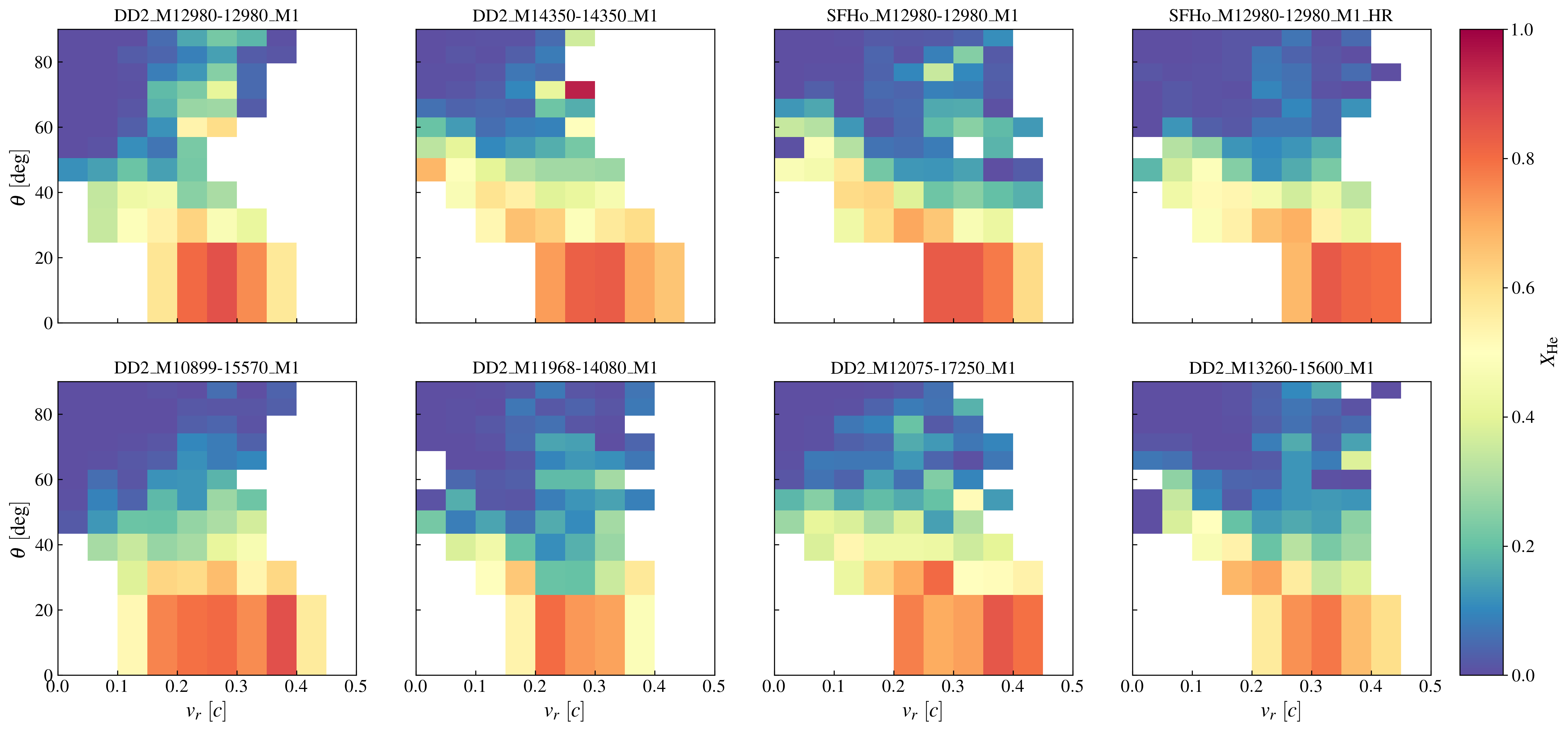}
    \caption{He mass fraction, $X_{\rm He}$, computed at $t = 1$ day as a function of expansion velocity, $v_r$, and polar angle, $\theta$, for all simulations listed in \reftab{tab:ejecta}. In each bin $(v_r,\theta)$ bin, $X_{\rm He}$ is computed as in \refeq{eq:XHe}.} 
    \label{fig:XHe}
\end{figure*}

The He and Sr mass fractions in \ac{BNS} merger ejecta have recently attracted considerable attention, owing to their impact on early kilonova spectra and their sensitivity to the lifetime of the merger remnant and the strength of neutrino-driven winds \citep[e.g.][]{Perego2022ApJ,Sneppen2026PhRvD,Chiba2026arXiv, Arya2026arXiv}. In particular, the abundances of He and Sr depend strongly on the thermodynamic properties of the ejecta, especially the electron fraction, making them valuable probes of the post-merger evolution. To facilitate future studies of the connection between ejecta coming from long-lived merger remnants and kilonova observables, we report the He and Sr mass fractions, $X_{\rm He}$ and $X_{\rm Sr}$ respectively, predicted by our simulations. 
In \reffig{fig:XHe}, we show $X_{\rm He}$ as a function of the radial velocity $v_r$ and polar angle $\theta$ at $t=1$ day after merger. 
As described in \refsec{sec:method_nuclear}, after the end of the hydrodynamical trajectory the network assumes homologous expansion at the velocity of the last hydrodynamical point. We therefore take $v_r$ of each tracer to be its velocity at the last hydrodynamical point, and $\theta$ to be its polar angle at the same point, assuming that both remain constant during the subsequent expansion.
We then construct a two-dimensional map of $X_{\rm He}(v_r,\theta)$ by binning all tracers on a grid of $10$ velocity bins, linearly spaced in $v_r\in[0,0.5]\,c$, and $11$ polar-angle bins, equally spaced in $\cos\theta$ between the pole and the equator. In each $(v_r,\theta)$ cell, we compute the $X_{\rm He}$ as
\begin{equation}\label{eq:XHe}
    X_{\rm He}(v_r,\theta) = \frac{\sum_i (X_{\rm He})_i\, m_i}{\sum_i m_i},
\end{equation}
where the sums run over all tracers $i$ in that cell, $m_i$ is the mass of tracer $i$, and $(X_{\rm He})_i$ is the single tracer He mass fraction. We find that $X_{\rm He}(v_r,\theta)$ follows a similar pattern in all our simulations: $X_{\rm He}$ is largest, reaching $0.6-0.8$, in the polar region at $v_r \gtrsim 0.2 \ c$, and decreases gradually toward lower velocities and the equatorial region, dropping below $0.1$ for $v_r < 0.2\,c$ near the equator. Notably, $X_{\rm He} > 0.1$ across the typical photospheric velocity range $0.2 \lesssim v_r/c \lesssim 0.3$ in all simulations, consistent with the predictions of \citet{Sneppen2026PhRvD} for mergers producing long-lived remnants. The one exception is the region very close to the equator ($\theta > 80^\circ$), where the He fraction falls below $0.1$ for some simulations.

\begin{figure}[htb!]
    \centering
    \includegraphics[scale = 0.8]{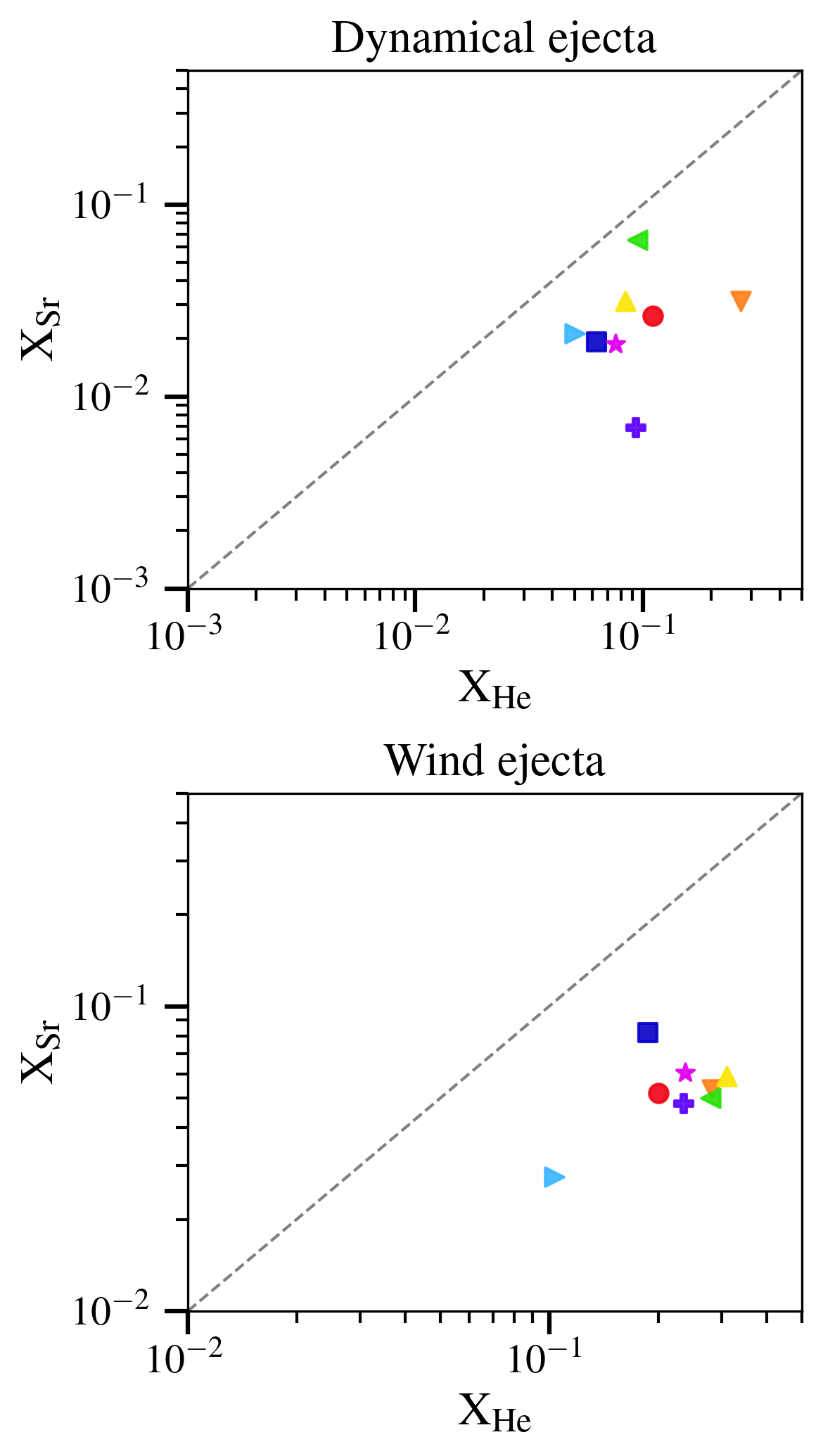}
    \caption{Sr mass fraction as a function of the He mass fraction for the dynamical ejecta (upper panel) and the wind ejecta (lower panel) in all simulations listed in \reftab{tab:ejecta}. The legend is the same as in \reffig{fig:ele_1day}.} 
    \label{fig:HeSr}
\end{figure}

In \reffig{fig:HeSr}, we compare the He and Sr mass fractions, $X_{\rm He}$ and $X_{\rm Sr}$, respectively, for the dynamical and wind ejecta in the $n = 0$ case. The mass fractions are defined as the mass of He or Sr synthesised in each ejecta component divided by the mass of the corresponding ejecta component. In all simulations, the He mass fraction exceeds that of Sr, $X_{\rm Sr} < X_{\rm He}$, in both ejecta components. The dynamical ejecta exhibit a relatively large model-to-model scatter. Therefore, for a fixed \ac{EOS}, the binary mass ratio and total mass have significant impact on the He and Sr mass fractions in the dynamical ejecta, with possible implications on kilonova spectral features sensitive to these quantities. On the contrary, the wind ejecta cluster around $X_{\rm He} \sim 0.2-0.3$ and $X_{\rm Sr} \sim 0.05$, with the exception of \texttt{DD2$\_$M10899-15570$\_$M1}, indicating a moderate dependence of these quantities on the \ac{BNS} system total mass and mass ratio. The total He mass fraction (dynamical plus wind) is at least $0.1$ in all simulations, with the minimum, $\simeq0.1$, reached by \texttt{DD2$\_$M10899-15570$\_$M1}.

\subsubsection{Yields at 1 Gyr}

\begin{figure*}[htb!]
    \centering
    \includegraphics[scale = 0.42]{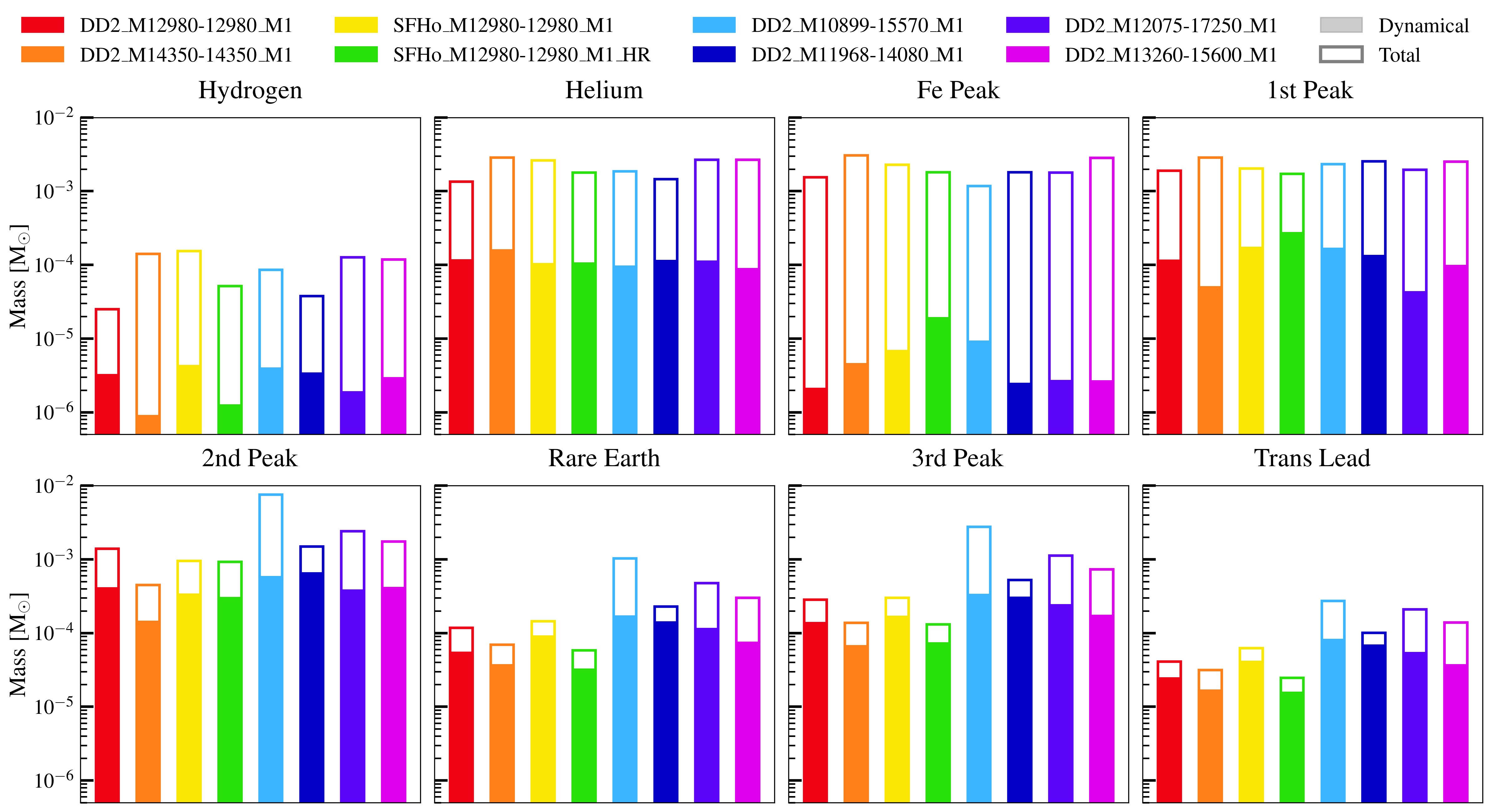}
    \caption{Nuclear yields computed at $t = 1$ Gyr for all simulations listed in \reftab{tab:bns_sims}. Open bars denote the total mass synthesised in each elemental group, while filled bars indicate the component from the dynamical ejecta. The Fe peak includes elements with $21 \leq Z \leq 30$. The first-peak, second-peak, rare-earth, third-peak, and trans-lead groups comprise isotopes with mass numbers $72 \leq A \leq 94$, $111 \leq A \leq 145$, $145 \leq A \leq 170$, $188 \leq A \leq 206$, and $A \geq 209$, respectively.}
    \label{fig:bar_plot}
\end{figure*}

In \reffig{fig:bar_plot}, we show the nuclear yields at $t = 1$ Gyr for all simulations listed in \reftab{tab:bns_sims}, grouped into Hydrogen (H), He, Fe-peak, first-peak, second-peak, rare-earth, third-peak, and trans-lead nuclei. Open bars denote the total mass synthesised in each group in the case $n=0$, while filled bars indicate the contribution of the dynamical ejecta alone. All models produce similarly large total masses ($ > 10^{-3}~M_\odot$) of He, Fe-peak, and first-peak nuclei, whereas the masses of second-peak, rare-earth, third-peak, and trans-lead nuclei vary substantially among models, particularly in the wind ejecta. The yields of H also depend strongly on the \ac{BNS} model, in both the dynamical and wind ejecta.

\begin{figure*}[htb!]
    \centering
    \includegraphics[scale = 1.5]{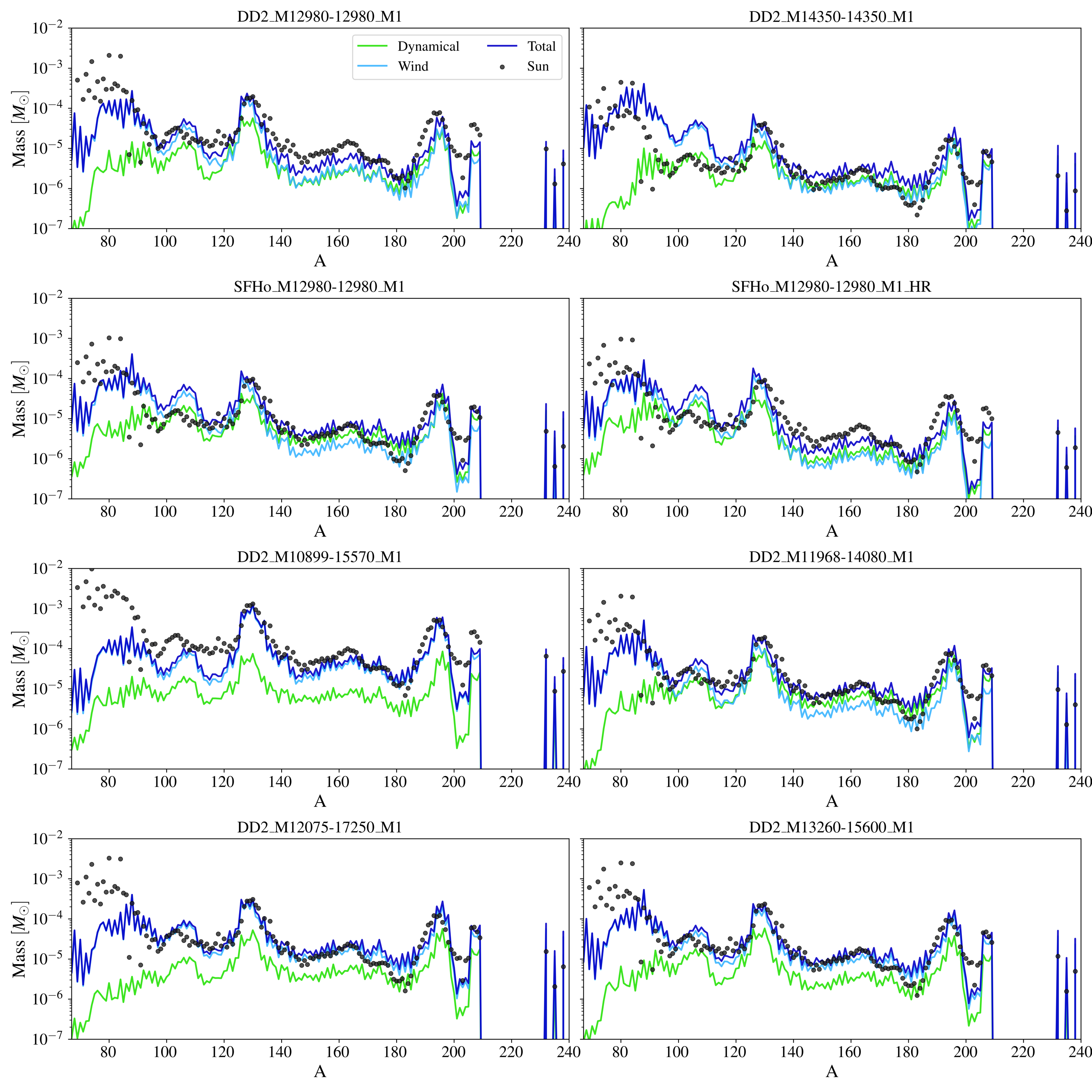}
    \caption{Final isotopic mass distribution for all simulations listed in \reftab{tab:ejecta}, considering the contribution from dynamical ejecta (green), wind ejecta (light blue), and the sum of the two (blue), compared to the solar $r$-process abundance (black dots) from \cite{Prantzos2020MNRAS}, normalized to $^{130}$Te.}
    \label{fig:solar_all}
    
\end{figure*}

The final abundance patterns are compared with the solar $r$-process residuals \citep{Prantzos2020MNRAS} in \reffig{fig:solar_all}, normalising the solar residuals to match the mass of the isotope $^{130}$Te in each simulation. We show in green and light blue the contributions from the dynamical and wind ejecta, respectively, and in blue the sum of the two. All simulations successfully reproduce the second and third $r$-process peaks, while exhibiting more pronounced differences in the first peak, the rare-earth (lanthanide) region, and the mass range between the first and second peaks. These variations primarily reflect the different amounts of low-$Y_e$ material ejected by the various merger models and the relative contributions of the dynamical and wind ejecta to the overall nucleosynthesis. For symmetric \ac{BNS} systems, the dynamical ejecta dominates the production of isotopes above the second peak ($A\gtrsim140$), while the wind ejecta dominates the production of isotopes below the second peak. The exception is \texttt{DD2$\_$M12980-12980$\_$M1}, for which the dynamical and wind ejecta contribute comparably above the second peak. For asymmetric mergers, instead, the largest contribution to the nucleosynthesis of all isotopes comes from the wind ejecta, since these systems have larger discs with a significant wind-ejecta component at $Y_e < 0.25$. An exception is \texttt{DD2$\_$M11968-14080$\_$M1}, which instead behaves like the symmetric simulations in terms of the division between dynamical and wind ejecta. This can be explained by the smaller wind ejecta of this simulation, whose wind phase lasts $5$--$10$~ms less than in the other asymmetric simulations.

Considering the total ejecta (blue curve), we see that while all our simulations successfully reproduce the second and third solar $r$-process peaks, they fail to reproduce the first peak. An exception is \texttt{DD2$\_$M14350-14350$\_$M1}, which reproduces the solar abundance reasonably well for $A<90$, but overproduces the $90<A<120$ and $170<A<190$ regions. Among our models, the equal-mass simulations \texttt{DD2$\_$M12980-12980$\_$M1} and \texttt{SFHo$\_$M12980-12980$\_$M1$\_$HR} provide the closest overall match to the solar $r$-process abundance pattern. The unequal-mass model \texttt{DD2$\_$M10899-15570$\_$M1} also reproduces the solar distribution remarkably well for nuclei beyond the second peak, although it significantly underproduces nuclei with $A \lesssim 120$. The remaining simulations generally fail to reproduce the solar abundances in the lanthanide region, overproducing this mass range relative to the observed pattern. All simulations tend to overproduce $^{232}$Th, $^{235}$U, and $^{238}$U relative to the solar $r$-process abundance pattern. However, we note that the abundances of isotopes from the second peak to the trans-lead region depend significantly on several uncertain nuclear inputs used in the $r$-process calculations (especially around $A \simeq 140$ and $A \simeq 203$), and are therefore themselves uncertain \citep{Eichler2019ApJ, Martinet2025A&A}.

\begin{figure*}[htb!]
    \centering
    \includegraphics[scale = 0.52]{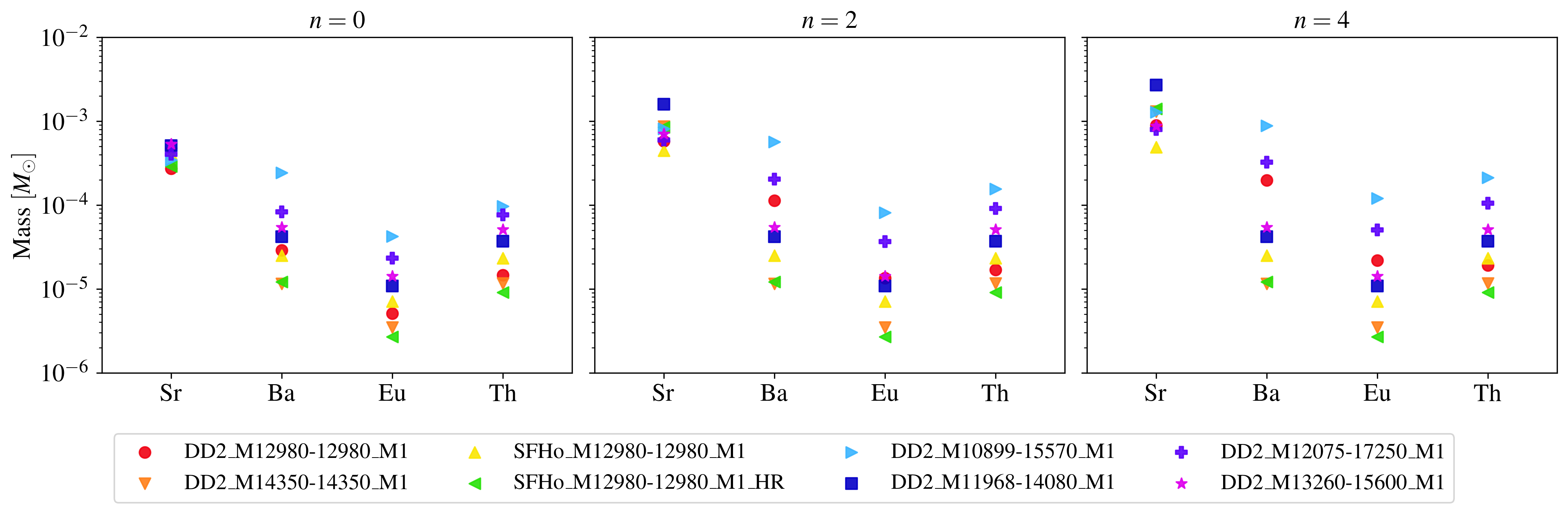}
    \caption{Masses of selected elements synthesised at $t = 1$ Gyr after merger for all simulations listed in \reftab{tab:ejecta}. From left to right, the wind ejecta mass is computed according to \refeq{eq:mwind}, by considering only the wind mass directly ejected during the simulation time ($n = 0$) and by adding $20\%$ or $40\%$ of the accretion disc mass ($n = 2$ and $n = 4$), respectively.}
    \label{fig:ele_1Gyr}
\end{figure*}

The differences among the nucleosynthesis yields can be quantified by comparing the masses of selected elements synthesised at $t = 1$ Gyr. In \reffig{fig:ele_1Gyr}, we show the total mass of Sr, Ba, Eu, and Th for all simulations in our sample, considering three prescriptions for the wind mass as in \refeq{eq:mwind}, that is $n = 0,~2,~4$. While all models predict similar Sr yields ($\sim 2 - 5 \times 10^{-4} M_\odot$), the synthesised masses of Ba, Eu, and Th vary by more than one order of magnitude across the simulation sample. 
Interestingly, in the first panel ($n = 0$), the equal-mass models \texttt{DD2$\_$M12980-12980$\_$M1} and \texttt{SFHo$\_$M12980-12980$\_$M1} produce very similar yields despite adopting different \acp{EOS}. By contrast, changing the masses of the two \acp{NS} while keeping the same \ac{EOS} results in a much larger variation in the predicted abundances. Furthermore, the numerical resolution also has a non-negligible impact: the difference between the \texttt{SFHo$\_$M12980-12980$\_$M1$\_$HR} and \texttt{SFHo$\_$M12980-12980$\_$M1} simulations is even larger than the effect of changing the \ac{EOS}. This suggests that numerical resolution can influence the predicted nucleosynthesis yields as strongly as the underlying microphysical model. 
As expected, increasing the wind contribution ($n > 0$), has a stronger impact on Sr production, increasing its production up to one order of magnitude, but a much more limited effect the production of Ba, Eu and Th, with the exception of simulations with asymmetric mass ratio, coherently with the discussion in \refsec{sec:yields_1d}.

\section{Discussion}\label{sec:discussion}

In this section, we compare our results to studies in which a different approach was used for the \ac{BNS} merger simulation, the neutrino treatment, or the nucleosynthesis calculations.

\citet{Kullman2022MNRAS} studied $r$-process nucleosynthesis in four \ac{BNS} merger models using two \acp{EOS}, DD2 and SFHo, each with a symmetric ($1.35$--$1.35\,M_\odot$) and an asymmetric ($1.25$--$1.45\,M_\odot$) configuration, all forming a long-lived \ac{NS} remnant surrounded by an accretion disc. Their simulations used a relativistic smoothed particle hydrodynamics (SPH) code \citep{Oechslin2007A&A, Bauswein2013ApJ}, with neutrino transport handled by the Improved Leakage-Equilibration-Absorption Scheme (ILEAS) \citep{Ardevol-Pulpillo2019MNRAS}, which refines standard leakage descriptions with a diffusion-time-scale prescription, a neutrino-equilibration treatment for the trapped regime, and a ray-tracing algorithm for neutrino re-absorption near the neutrinosphere. They identified dynamical ejecta as all mass elements located beyond $100$~km at $7$~ms after merger whose kinetic plus thermal energy exceeds their gravitational binding energy. We notice that the nuclear inputs used in their nucleosynthesis calculations differ substantially from those adopted here, and refer the reader to \citet{Kullman2022MNRAS} for details.
A comparison of the properties of the dynamical ejecta and the associated nucleosynthesis between our simulations and those of \citet{Kullman2022MNRAS} shows good overall agreement. Specifically, the mass of dynamically ejected matter in our simulations ranges over $[0.60, 1.77]\times10^{-3}\,M_\odot$, lower than but still comparable to their range of $[0.87, 2.56]\times10^{-3}\,M_\odot$. The mass-weighted average electron fraction used as input for the nucleosynthesis calculations (i.e. $Y_e$ evaluated once $T$ is permanently below $7$~GK) ranges over $\langle Y_e\rangle = 0.17$--$0.30$ in our simulations, in good agreement with the range $\langle Y_e\rangle = 0.22$--$0.27$ reported by \citet{Kullman2022MNRAS}. Our lower values of $\langle Y_e\rangle$ arise because our sample includes very asymmetric systems, which tend to produce more neutron-rich dynamical ejecta.
Even though we adopt different nuclear inputs, the mass fractions of lanthanides plus actinides ($X_{\rm LA}$), of $r$-process nuclei with $A>69$ ($X_{A>69}$), and of third-peak $r$-process nuclei with $A>183$ ($X_{A>183}$) in the dynamical ejecta are all compatible with \citet{Kullman2022MNRAS}: we find $X_{\rm LA} = 0.054-0.184$, compared to their $X_{\rm LA} = 0.107-0.152$; $X_{A>69} = 0.72-0.94$, compared to their $X_{A>69} = 0.88-0.95$; and $X_{A>183} = 0.09-0.30$, compared to their $X_{A>183} = 0.11-0.15$. Our results show a larger spread than \citet{Kullman2022MNRAS}, reflecting the broader diversity of masses and mass ratios explored in our simulations: the largest values are found in our most asymmetric binaries, while the smallest come from the equal-mass SFHo (HR) and the equal-mass DD2 model with $M=1.298\,M_\odot$ per star.

In a follow-up work, \citet{Kullmann2023MNRAS} explored the impact of nuclear uncertainties on the nucleosynthesis of \ac{BNS} merger ejecta by varying the nuclear mass model, $\beta$-decay rates, and fission properties. In particular, they computed the Th/U ratio for three simulations performed with the SFHo \ac{EOS} and binary masses of $1.25$--$1.45$ and $1.35$--$1.35\,M_\odot$, modelling the post-merger remnant as a central black hole of $3\,M_\odot$ surrounded by a disc of $0.3\,M_\odot$, and varying 15 possible nuclear input configurations. They computed $\log\epsilon(\rm{Th/U})$ from their simulation output, considering only the $^{232}$Th and $^{238}$U isotopes. They found that, while variations in the nuclear input produce a significant spread in the Th/U ratio, changing the binary system has comparatively little effect on the result. Specifically, they found $\log\epsilon(\rm{Th/U}) = 0.04$--$0.31$ for the $1.25$--$1.45\,M_\odot$ system and $\log\epsilon(\rm{Th/U}) = 0.07$--$0.37$ for the $1.35$--$1.35\,M_\odot$ system at $t = 1$ Gyr. We find a fully compatible, but much narrower, range of $0.20$--$0.23$, since we vary only the \ac{EOS} and mass ratio, and not the nuclear inputs.

\citet{Fujibayashi2020ApJ} studied the nucleosynthesis of post-merger ejecta from low-mass \ac{NS} binaries, including two equal-mass \ac{BNS} systems with the DD2 \ac{EOS}: one with $M_1=M_2=1.25\,M_\odot$, and the other with $M_1=M_2=1.35\,M_\odot$. Their fully general-relativistic simulations included neutrino radiation and viscous hydrodynamics, using an M1 closure scheme for free-streaming neutrinos and a leakage scheme for trapped neutrinos. Both binaries were evolved for $\sim6$~s after merger, with $\sim40\,000$ Lagrangian tracers per simulation used for nucleosynthesis calculations. 
Looking at the nucleosynthesis results from these two models, computed using an $\alpha$-viscosity parameter of $0.04$, they found a solar $r$-process-like abundance pattern over a wide range of $A$ ($\sim80$--$200$) only when the post-merger elapsed time was below $0.2$~s. The subsequent mass ejection added a substantial amount of matter with $A<130$, so that the abundances of the heavier nuclei were lower than the solar $r$-process residuals. The resulting overall patterns were instead consistent with that of the $r$-process deficient metal-poor star HD 122563 \citep{Cowan2005ApJ, Honda2006ApJ, Roederer2012ApJS}.
Comparing with our simulation \texttt{DD2$\_$M12980-12980$\_$M1}, whose total mass lies between those of the two \citet{Fujibayashi2020ApJ} models, we find that, for $n=0$, we reproduce the solar $r$-process abundance well except for the first peak, as shown in  \reffig{fig:solar_all}. Adopting instead the $n=4$ prescription -- consistent with a central remnant surviving beyond the end of our simulation -- we reproduce the solar $r$-process pattern well up to the second peak, but significantly underproduce isotopes with $A>140$.

\citet{Just2023ApJ} investigated the nucleosynthesis in \ac{BNS} merger models with remnant lifetimes of $\sim0.1$--$1$~s before \ac{BH} formation, following the dynamical ejecta, the ejecta launched during the \ac{NS}-remnant phase (NS-torus ejecta), and the ejecta from the viscous disintegration of the \ac{BH} torus (BH-torus ejecta). The evolution was followed up to $10$~ms after merger with a 3D general-relativistic SPH code and the ILEAS neutrino scheme \citep{Ardevol-Pulpillo2019MNRAS}. The subsequent post-merger evolution was axisymmetric, performed with the special-relativistic code ALCAR-AENUS \citep{Just2015MNRAS} with energy-dependent M1 neutrino transport, and the ejecta were followed up to $100$~s. The post-merger evolution extended the classical $\alpha$-viscosity scheme to capture MRI-driven viscosity in both the accretion torus and the \ac{NS} remnant, with a dedicated parameter, $n_{\rm vis}$, controlling the remnant viscosity independently of the disc. Tracers were sampled from the final homologous ejecta, with $4000$--$5000$ effective tracers per model for the dynamical component, and their histories were reconstructed by backward integration in time. Nucleosynthesis was computed with two independent nuclear networks. Their sample comprised ten models, with symmetric ($q=1$) and asymmetric ($q=0.75$) binaries of total mass $2.75\,M_\odot$, all with the SFHo \ac{EOS}, for which they varied $n_{\rm vis}$ and the $\alpha$-viscosity parameter $\alpha_{\rm vis}$. \citet{Just2023ApJ} summarised in their Table~1 the properties of the dynamical, \ac{NS}-torus and \ac{BH}-torus ejecta. Since the \ac{NS}-torus component included the neutrino-driven wind, we compare their dynamical and \ac{NS}-torus ejecta with our dynamical and wind ejecta, respectively. Because the two studies define the ejecta components differently, this comparison is only approximate.
Their dynamical ejecta have masses of $6$--$13\times10^{-3}\,M_\odot$, larger than ours ($0.6$--$1.8\times10^{-3}\,M_\odot$), a difference that is probably due, at least in part, to the different definitions of dynamical ejecta. For $n=0$, our wind ejecta ($6.0$--$16.9\times10^{-3}\,M_\odot$) are also systematically smaller than their \ac{NS}-torus ejecta ($18$--$39\times10^{-3}\,M_\odot$). This is expected, since our simulations end before the collapse of the remnant to a \ac{BH} in their models.
The dynamical ejecta of our simulations span a broader range of electron fractions. Their average $Y_e$ (measured at $5$~GK) lies in $0.23$--$0.25$, while we find $\langle Y_e\rangle=0.17$--$0.30$ (also evaluated at $5$~GK, for consistency with their definition). This reflects the larger variety of total binary masses, mass ratios and \acp{EOS} in our sample, against a single \ac{EOS} and a fixed total mass in theirs. The lanthanide-plus-actinide mass fraction follows the same trend: they find $X_{\rm LA}=0.118$--$0.147$, whereas we find $X_{\rm LA}=0.054$--$0.184$. Our lowest value comes from \texttt{SFHo$\_$M12980-12980$\_$M1$\_$HR}, whose dynamical ejecta have $\langle Y_e\rangle \simeq 0.3$.
For the \ac{NS}-torus ejecta, the average electron fractions are similar ($0.37$--$0.42$ for them, $0.35$--$0.43$ for us). The lanthanide-plus-actinide fraction, however, is much smaller in their models ($X_{\rm LA}\lesssim4.4\times10^{-3}$) than in ours ($6.0$--$58.7\times10^{-3}$ for $n=0$). Two effects probably contribute: our wind ejecta may include material that they classify as dynamical, and in most of their models the \ac{NS} survives for more than $100$~ms, which increases the contribution of the neutrino-driven wind and lowers $X_{\rm LA}$. In agreement with this picture, for $n=4$, which gives more weight to the neutrino-driven wind, our $X_{\rm LA}$ falls to $0.7$--$38.3\times10^{-3}$, closer to their range. A comparison of the final nucleosynthesis yields is not straightforward, because our simulations have no \ac{BH}-torus ejecta. However, it is worth to note that our simulations predict similar He mass fraction.

\citet{Ricigliano2024MNRAS2} investigated the impact of the nuclear \ac{EOS} on nucleosynthesis by simulating symmetric \ac{BNS} mergers with \ac{NS} masses of $1.365 \ M_\odot$, using eight different
finite-temperature nuclear \acp{EOS}. The latter were obtained by systematically varying nuclear properties, such as the incompressibility and the nucleon effective mass at saturation density,
in a Skyrme-type interaction model. 
Our approach differs from theirs mainly in two respects: the tracer method and the neutrino transport scheme. We evolve a large number ($2^{18}$) of passive tracers,  initialised at the beginning of the simulation, select those that are ejected, and subsample $3000-5000$ representative tracers per simulation, as described in Sects. \ref{subsec:ejected_matter} --\ref{subsec:trcrs_part}. \citet{Ricigliano2024MNRAS2}, instead, placed $\sim 15\,000$ tracers in the ejecta at the end of each 3D simulation and integrated their trajectories backward in time. 
Moreover, they adopted an M0 neutrino transport scheme, while we use the more refined M1 scheme. Despite these differences, we find comparable dynamical and wind ejecta masses; the variability in our sample stems mainly from differences in mass ratio, whereas in theirs it stems from the \ac{EOS}, suggesting that varying the mass ratio and varying the \ac{EOS} can produce similar levels of variability in the ejecta mass.
The properties of the ejected matter are also broadly similar, though with some notable differences. In both our dynamical and wind ejecta, we find a significant component with $Y_e > 0.4$ that is absent in \citet{Ricigliano2024MNRAS2}; in the wind ejecta, we also find more mass in the polar region than they reported. Both differences are consistent with the M0 versus M1 neutrino treatment. The resulting nucleosynthesis is likewise similar overall, with a few  exceptions.
We find that our total He yield computed at one day is significantly larger than that reported by \cite{Ricigliano2024MNRAS2}, even for the $n=0$ case.  
Specifically, the He mass synthesised in the dynamical ejecta exceeds their values by a factor of approximately $2\text{--}5$, while the wind ejecta produce an order of magnitude more He. Comparing our \reffig{fig:bar_plot} with figure~7 of \citet{Ricigliano2024MNRAS2}, we find that, while the H and He mass fraction from the dynamical ejecta is comparable between the two studies, the H and He mass fraction from the wind/disc ejecta is an order of magnitude larger in our case, for all our simulations. This difference is consistent with the different neutrino transport schemes adopted in the two studies. The stronger neutrino irradiation in the M1 treatment leads to more proton-rich ejecta, thereby favouring the synthesis of light nuclei such as H, He and Fe-peak nuclei. By comparing the final nucleosynthesis at 1 Gyr, we find that even though we get comparable Sr yields, our simulations systematically produce larger amounts of Eu and Th than \cite{Ricigliano2024MNRAS2}, in some cases by up to an order of magnitude. This enhancement is particularly pronounced for the unequal-mass systems. Two effects likely contribute to this difference. First, the asymmetric mergers eject a larger fraction of neutron-rich material, favouring the synthesis of nuclei beyond the second $r$-process peak. Second, differences in the adopted nuclear physics inputs, particularly in the calculation of the nuclear reaction rates and fission fragments, can significantly affect the production of lanthanides and actinides \citep{Eichler2019ApJ, Martinet2025A&A}. A systematic investigation of the impact of the nuclear physics inputs on our yields is left to future work.

\section{Nuclear yields in the \texttt{MARTINI} platform}\label{sec:Martini}

\texttt{MARTINI}\footnote{\url{martini.oa-abruzzo.inaf.it}} is an online platform dedicated to stellar nucleosynthesis and the study of kilonova light curves and spectra (\citealt{Bezmalinovich2026A&A}). The platform already hosts nucleosynthesis yields from the slow neutron capture process ($s$-process), previously released through the FRUITY database (\citealt{Cristallo2011ApJS,Cristallo2015ApJS}), together with atomic data for Se and Zr (\citealt{Bezmalinovich2026A&A, Bezmalinovich2026Galaxies}). 

\begin{figure*}[htb!]
    \centering
    \includegraphics[scale = 0.37]{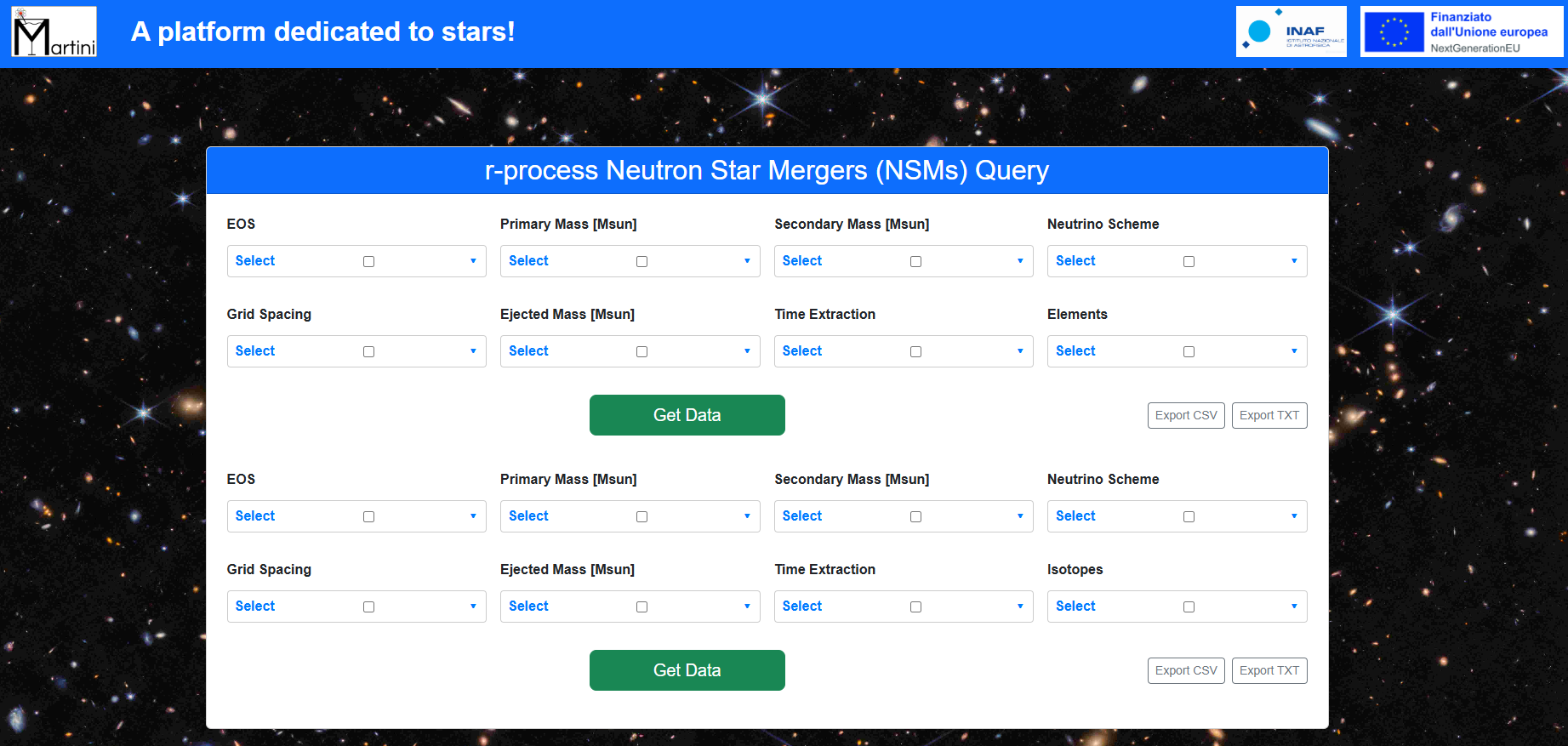}
    \caption{Query form for NSM r-process yields available in the \texttt{MARTINI} platform.}
    \label{fig:martini}
\end{figure*}

The $r$-process yields from \ac{BNS} mergers presented in this article are now also available through \texttt{MARTINI}. The platform allows users to visualize the yields online and to download them in \texttt{.txt} or \texttt{.csv} formats. The query interface, shown in Fig. \ref{fig:martini}, enables users to retrieve elemental or isotopic yields for different combinations of binary masses, \ac{EOS}, numerical resolution, neutrino transport scheme, and wind ejecta mass prescription (cases $n = 0 - 4$ as in \refeq{eq:mwind}). The published yields are provided at $t = 1$ day after merger, relevant for kilonova studies, and at $t = 1$ Gyr, relevant for studies of galactic chemical evolution. Users interested in the set of tracers used to compute the nucleosynthesis and the final yields can download them from \url{https://doi.org/10.5281/zenodo.23689687} . 

In the following section, we discuss an example of usage of the yields available through the \texttt{MARTINI} platform.

\subsection{Example of application: chronometer pairs}

The nucleosynthesis yields released through the \texttt{MARTINI} platform can be readily evolved beyond $t = 1$ Gyr to compute radioactive chronometer pairs, such as Th/Eu, for merger models spanning a wide range of ejecta properties and $r$-process outcomes. These predictions can be directly compared with abundance measurements in metal-poor and actinide-boost stars, providing a valuable tool for investigating the origin of the observed actinide enhancement. In particular, they contribute to the long-standing debate over whether actinide-boost stars can be explained by different ejecta conditions within a single $r$-process site, such as \ac{BNS} mergers, or whether they instead require the existence of multiple distinct astrophysical $r$-process sites \citep{Eichler2019ApJ, Holmbeck2019ApJ}. 

\begin{figure*}[htb!]
    \centering
    \includegraphics[scale = 0.55]{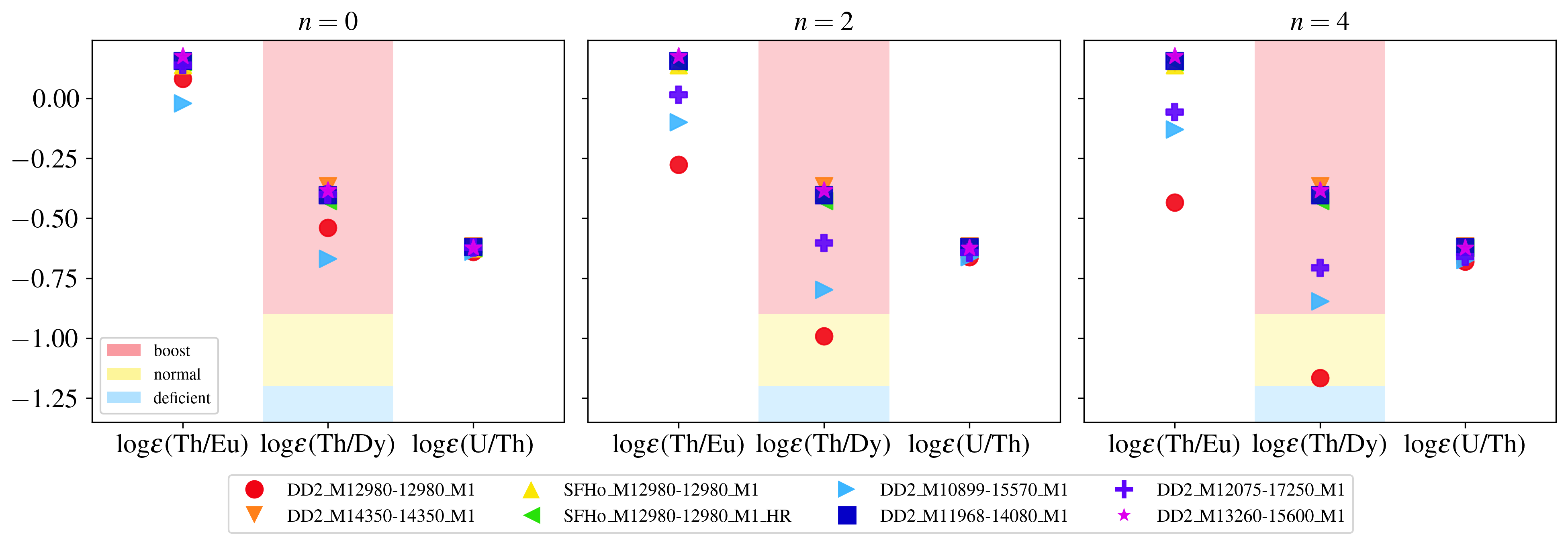}
    \caption{$\log \epsilon (\rm{Th/Eu})$, $\log \epsilon (\rm{Th/Dy})$, and $\log \epsilon (\rm{U/Th})$ computed at $t=10$ Gyr for all simulations listed in \reftab{tab:bns_sims}. From left to right, the wind ejecta mass is computed according to \refeq{eq:mwind}, by considering only the wind mass directly ejected during the simulation time ($n = 0$) and by adding $20\%$ or $40\%$ of the accretion disc mass ($n = 2$ and $n = 4$), respectively. The three shaded regions mark the classification scheme of $r$-process stars as reported in \citet{Holmbeck2019ApJ}: actinide-deficient (blue), actinide-normal (yellow), and actinide-boost (pink).}
    \label{fig:ratio_10Gyr}
\end{figure*}

To compute the chronometer pairs Th/Eu, Th/Dy, and U/Th at $t = 10$ Gyr, it is sufficient to evolve the abundances released at $t = 1$ Gyr over an additional 9 Gyr, taking into account the radioactive decay of $^{232}\text{Th}$ and $^{238}\text{U}$. 
The abundance ratios of the pairs Th/Eu, Th/Dy, and U/Th, computed at $t=10$ Gyr, are shown in \reffig{fig:ratio_10Gyr}. The wind ejecta mass is calculated according to \refeq{eq:mwind}, considering either only the material directly ejected during the simulation ($n = 0$) or including an additional $20\%$ or $40\%$ of the accretion disc mass ($n = 2$ and $n = 4$, respectively). Variations in both the \ac{BNS} model and the assumed wind ejecta mass have a significant impact on $\log \epsilon (\rm{Th/Eu})$ and $\log \epsilon (\rm{Th/Dy})$. This behaviour is consistent with the findings of \citet{Holmbeck2019ApJ}, who showed that that $\log \epsilon (\rm{Th/Dy})$ is extremely sensitive to the ejecta $Y_e$ and entropy. \citet{Holmbeck2019ApJ} define $r$-process stars as actinide-deficient if $\log\epsilon(\rm{Th/Dy}) < -1.2$, actinide-normal if $-1.2 < \log\epsilon(\rm{Th/Dy}) < -0.9$, and actinide-boost if $\log\epsilon(\rm{Th/Dy}) > -0.9$. We report these three groups in \reffig{fig:ratio_10Gyr} as blue, yellow, and pink shaded regions, respectively.

For the $n=0$ case, the Th/Dy ratios predicted by all our simulations fall in the actinide-boost region. In particular, the \texttt{DD2$\_$M10899-15570$\_$M1} and \texttt{DD2$\_$M12980-12980$\_$M1} simulations fall within the range $[-0.9, -0.6]$ observed in actinide-boost stars in the Milky Way \citep[e.g.][figure~1]{Holmbeck2019ApJ}. When the wind contribution is larger ($n>0$), \texttt{DD2$\_$M12980-12980$\_$M1} becomes consistent with the Th/Dy ratios measured in actinide-normal metal-poor Milky Way stars, and approaches the actinide-deficient range for $n=4$. The models \texttt{DD2$\_$M10899-15570$\_$M1} and \texttt{DD2$\_$M12075-17250$\_$M1} remain in the actinide-boost region for $n=2$ and start approaching the actinide-normal region for $n=4$. All other models keep $\log\epsilon(\rm{Th/Dy})\sim-0.4$, consistent with actinide-boost stars, for all $n>0$.

Considering the Th/Eu ratio, we find that for $n=0$ all models fall in the actinide-normal or actinide-boost range of the \citet{Holmbeck2019ApJ} classification. Only \texttt{DD2$\_$M10899-15570$\_$M1} is marginally consistent with the metal-poor star measurements of \citet{Roederer2009ApJ}. When the wind contribution is larger ($n>0$), \texttt{DD2$\_$M12980-12980$\_$M1} and \texttt{DD2$\_$M12075-17250$\_$M1} also become consistent with the observed stellar abundances \citep{Roederer2009ApJ}.

As expected, $\log \epsilon (\rm{U/Th})$ is largely insensitive to both the \ac{BNS} model and wind ejecta mass \citep[e.g.][]{Kullman2022MNRAS}. All models predict $\log \epsilon (\rm{U/Th}) \sim -0.6$, in good agreement with measurements in metal poor stars \cite{Roederer2009ApJ}.

However, we note that all of these abundance ratios are highly sensitive to the nuclear inputs adopted in the nucleosynthesis calculations \citep[e.g.][]{Eichler2019ApJ, Kullmann2023MNRAS}. This additional source of theoretical uncertainty should therefore be carefully considered when comparing nucleosynthesis predictions with observational data.

\section{Conclusions}\label{sec:conclusions}
In this work, we have developed and publicly released a new set of $r$-process nucleosynthesis yields based on a sample of eight \ac{BNS} merger simulations in numerical relativity. To our knowledge, this is the first public release of yields and tracer trajectories obtained from full general relativity simulations of long-lived remnants with M1 neutrino transport.

Starting from a sample of eight \ac{BNS} merger models spanning different \acp{EOS}, total binary masses, and mass ratios, we performed numerical relativity simulations and recorded both the thermodynamic properties of the ejecta crossing the detector surface and the thermodynamic evolution of passive tracer particles. All simulations employed the M1 neutrino transport scheme and resulted in long-lived \ac{NS} remnants. We then developed a sampling algorithm to select $3000-5000$ representative tracers per simulation and assigned a representative mass to each tracer. The resulting tracer set was post-processed with the nuclear reaction network  \texttt{WinNet} to compute $r$-process nucleosynthesis yields. Finally, we released the nucleosynthesis yields for all simulations, computed at $t = 1$ day and $t = 1$ Gyr, considering multiple prescriptions for the wind ejecta mass to account for the current uncertainties in the amount of material unbound from the remnant accretion disc.
The main results of this study can be summarised as follows.

We publicly release the complete set of tracer trajectories for all \ac{BNS} merger models together with the corresponding nucleosynthesis yields through the online platform \texttt{MARTINI}. The selected tracers accurately reproduce the thermodynamic and mass distributions of both the dynamical and wind ejecta, while the published yields at $t=1$ day and $t=1$ Gyr provide a resource for future kilonova and Galactic chemical evolution studies.

The nucleosynthesis yields exhibit a clear dependence on the merger properties. At $t = 1$ day, while the production of light and first-peak elements is relatively robust across the simulation sample and is primarily enhanced by the neutrino-irradiated disc wind, the yields of heavy $r$-process nuclei vary by more than one order of magnitude, with unequal-mass mergers producing the largest abundances owing to their larger fraction of neutron-rich ejecta. Compared with previous calculations employing an M0 neutrino transport scheme, our M1 simulations predict substantially larger He abundances, particularly in the wind ejecta, demonstrating the important role of neutrino irradiation in determining the ejecta composition and the resulting nucleosynthesis.

In all simulations, the total He mass fraction of the ejecta (dynamical plus wind, for $n=0$) exceeds $0.1$, and $X_{\rm He}>X_{\rm Sr}$ in both the dynamical and wind ejecta. At $t = 1$ day, the wind ejecta cluster around $X_{\rm He}\sim0.2-0.3$ and $X_{\rm Sr}\sim0.05$ for all our simulations with the exception of \texttt{DD2$\_$M10899-15570$\_$M1}.

At $t=1$ Gyr, all simulations consistently produce large and comparable yields of light, Fe-peak, and first-peak elements, while the yields of second-peak, rare-earth, and third-peak nuclei vary substantially among models, particularly in the wind ejecta, with asymmetric \ac{BNS} models generally producing more rare-earth and third-peak elements. In particular, the production of first-peak elements such as Sr is remarkably robust among simulations, whereas the yields of heavier elements, including Ba, Eu, and Th, span more than one order of magnitude. These variations primarily reflect the different amounts of neutron-rich ejecta and the relative contributions of the dynamical and wind components.
All models reproduce the second and third solar $r$-process peaks, but all except \texttt{DD2$\_$M14350-14350$\_$M1} fail to reproduce the first one, and several of them overproduce the rare-earth region. The models giving the closest match to the solar abundance pattern are the equal-mass \texttt{DD2$\_$M12980-12980$\_$M1} and \texttt{SFHo$\_$M12980-12980$\_$M1$\_$HR}.

The nucleosynthesis yields presented in this work are, however, subject to several sources of uncertainty. From the nuclear physics side, the predicted abundances depend sensitively on the adopted nuclear mass model and reaction rates \citep[e.g.][]{Eichler2019ApJ}. From the numerical relativity side, we have shown that the simulation resolution can influence the yields, in some cases at a level comparable to that of changing the \ac{NS} \ac{EOS}. In addition, the ejecta mass had not saturated at the end of our simulations, so the wind ejecta masses $m_{\rm wind}^{n = 0}$ should be regarded as lower limits. The thermodynamic evolution of the ejecta is also influenced by the uncertainties in the modelling of the merger microphysics, and the treatment of viscosity can significantly affect the nucleosynthesis output \citep{Fujibayashi2020ApJ, Just2023ApJ}.
A further source of uncertainty is the assumption of homologous expansion when post-processing tracers for the nucleosynthesis calculations \citep{Magistrelli2026A&A}. Moreover, the present dataset is limited to \ac{BNS} mergers producing long-lived remnants, and should be extended in the future to include systems undergoing prompt or delayed collapse to a \ac{BH}. Finally, it will be important to assess the robustness of our results against the adopted tracer mass-assignment procedure, by comparing with state-of-the-art prescriptions in which tracers are sampled from the ejecta at the detector sphere, and their properties reconstructed by integrating their trajectories backward in time \citep[e.g.][]{Nishimura2015ApJ, Fujibayashi2020ApJ, Witt2021ApJ, Just2023ApJ, Jacobi2024MNRAS}.

Despite these uncertainties, the dataset presented here provides the first publicly available library of $r$-process yields and tracer trajectories extracted from \ac{BNS} merger simulations performed with the M1 neutrino transport scheme, offering a valuable resource for future studies of kilonovae, Galactic chemical evolution, and stellar abundances.

\section*{Data availability}
The tracer trajectories (time, temperature, density, radius, and electron fraction) and the tracer masses for all the simulations presented in this work are publicly available on Zenodo at \url{https://doi.org/10.5281/zenodo.23689687}. The nucleosynthesis yields, both elemental and isotopic, at 1 day and 1 Gyr are available through the \texttt{MARTINI} platform at \url{https://martini.oa-abruzzo.inaf.it}.
\begin{acknowledgements}
      E.L., D.V., and S.C. acknowledge funding by the European Union – NextGenerationEU RFF M4C2 1.1 PRIN 2022 project 2022RJLWHN URKA. E.L. and S.C. acknowledge founding from 1.05.24.07.02 MiniGrants RSN2 (2024). 
      A.P. and F.M.G. are supported by the European Union under NextGenerationEU, PRIN
2022 Project No. 2022KX2Z3B.
      A.P. and F.M.G. also acknowledge the EuroHPC Joint Undertaking for awarding this project access to the EuroHPC supercomputer LUMI, hosted by CSC (Finland) and the LUMI consortium through a EuroHPC Extreme Scale Access call (EHPC-EXT-2022E01-046).
      The authors acknowledge 
      computational resources from the INFN-Torino (Tullio server) and INFN-TIFPA (TheoServ server).
      The authors thank Jan Kuske, Almudena Arcones, Gabriele Cescutti, Marta Molero, and Giacomo Ricigliano for useful interactions and comments.
\end{acknowledgements}

\bibliographystyle{bibtex/aa}
\bibliography{bibtex/biblio}

\providecommand{\noopsort}[1]{}\providecommand{\singleletter}[1]{#1}%
\begin{thebibliography}{122}
\expandafter\ifx\csname natexlab\endcsname\relax\def\natexlab#1{#1}\fi

\bibitem[{{Abbott} {et~al.}(2017{\natexlab{a}}){Abbott}, {Abbott}, {Abbott}, {Acernese}, {Ackley}, {Adams}, {Adams}, {Addesso}, {Adhikari}, {Adya}, {Affeldt}, {Afrough}, {Agarwal}, {Agathos}, {Agatsuma}, \& {Aggarwal}}]{Abbott2017PhRvLgw170817}
{Abbott}, B.~P., {Abbott}, R., {Abbott}, T.~D., {et~al.} 2017{\natexlab{a}}, \prl, 119, 161101

\bibitem[{{Abbott} {et~al.}(2017{\natexlab{b}}){Abbott}, {Abbott}, {Abbott}, {Acernese}, {Ackley}, {Adams}, {Adams}, {Addesso}, {Adhikari}, {Adya}, {Affeldt}, {Afrough}, {Agarwal}, {Agathos}, {Agatsuma}, \& {Aggarwal}}]{Abbott2017ApJgw170817}
{Abbott}, B.~P., {Abbott}, R., {Abbott}, T.~D., {et~al.} 2017{\natexlab{b}}, \apjl, 848, L12

\bibitem[{{Antoniadis} {et~al.}(2013){Antoniadis}, {Freire}, {Wex}, {Tauris}, {Lynch}, {van Kerkwijk}, {Kramer}, {Bassa}, {Dhillon}, {Driebe}, {Hessels}, {Kaspi}, {Kondratiev}, {Langer}, {Marsh}, {McLaughlin}, {Pennucci}, {Ransom}, {Stairs}, {van Leeuwen}, {Verbiest}, \& {Whelan}}]{Antoniadis2013Science}
{Antoniadis}, J., {Freire}, P. C.~C., {Wex}, N., {et~al.} 2013, Science, 340, 448

\bibitem[{{Ardevol-Pulpillo} {et~al.}(2019){Ardevol-Pulpillo}, {Janka}, {Just}, \& {Bauswein}}]{Ardevol-Pulpillo2019MNRAS}
{Ardevol-Pulpillo}, R., {Janka}, H.-T., {Just}, O., \& {Bauswein}, A. 2019, \mnras, 485, 4754

\bibitem[{{Arya} {et~al.}(2026){Arya}, {Damgaard}, {Sneppen}, {Dougan}, {Sim}, {Ballance}, \& {Watson}}]{Arya2026arXiv}
{Arya}, A., {Damgaard}, R., {Sneppen}, A., {et~al.} 2026, arXiv e-prints, arXiv:2604.05812

\bibitem[{{Bauswein} {et~al.}(2013){Bauswein}, {Goriely}, \& {Janka}}]{Bauswein2013ApJ}
{Bauswein}, A., {Goriely}, S., \& {Janka}, H.~T. 2013, \apj, 773, 78

\bibitem[{{Bernuzzi} \& {Hilditch}(2010)}]{Bernuzzi2010PrD}
{Bernuzzi}, S. \& {Hilditch}, D. 2010, \prd, 81, 084003

\bibitem[{{Bernuzzi} {et~al.}(2025){Bernuzzi}, {Magistrelli}, {Jacobi}, {Logoteta}, {Perego}, \& {Radice}}]{Bernuzzi2025MNRAS}
{Bernuzzi}, S., {Magistrelli}, F., {Jacobi}, M., {et~al.} 2025, \mnras, 542, 256

\bibitem[{{Bezmalinovich}(2026)}]{Bezmalinovich2026Galaxies}
{Bezmalinovich}, M. 2026, Galaxies, 14, 62

\bibitem[{{Bezmalinovich} {et~al.}(2026){Bezmalinovich}, {Bulla}, {Gaigalas}, {Vescovi}, {Canzari}, \& {Cristallo}}]{Bezmalinovich2026A&A}
{Bezmalinovich}, M., {Bulla}, M., {Gaigalas}, G., {et~al.} 2026, \aap, 708, A366

\bibitem[{{Burbidge} {et~al.}(1957){Burbidge}, {Burbidge}, {Fowler}, \& {Hoyle}}]{Burbidge1957RvMP}
{Burbidge}, E.~M., {Burbidge}, G.~R., {Fowler}, W.~A., \& {Hoyle}, F. 1957, Reviews of Modern Physics, 29, 547

\bibitem[{{{\c{C}}okluk} {et~al.}(2026){{\c{C}}okluk}, {Perego}, {Guercilena}, \& {Yakut}}]{Cokluk2026arXiv260816275C}
{{\c{C}}okluk}, K.~A., {Perego}, A., {Guercilena}, F.~M., \& {Yakut}, K. 2026, arXiv e-prints, arXiv:2608.16275

\bibitem[{{Chiba} {et~al.}(2026){Chiba}, {Tanaka}, {Wanajo}, {Fujibayashi}, {Kawaguchi}, \& {Hotokezaka}}]{Chiba2026arXiv}
{Chiba}, K., {Tanaka}, M., {Wanajo}, S., {et~al.} 2026, arXiv e-prints, arXiv:2604.05703

\bibitem[{{Cowan} {et~al.}(2005){Cowan}, {Sneden}, {Beers}, {Lawler}, {Simmerer}, {Truran}, {Primas}, {Collier}, \& {Burles}}]{Cowan2005ApJ}
{Cowan}, J.~J., {Sneden}, C., {Beers}, T.~C., {et~al.} 2005, \apj, 627, 238

\bibitem[{{Cowan} {et~al.}(2021){Cowan}, {Sneden}, {Lawler}, {Aprahamian}, {Wiescher}, {Langanke}, {Mart{\'\i}nez-Pinedo}, \& {Thielemann}}]{Cowan2021RvMP}
{Cowan}, J.~J., {Sneden}, C., {Lawler}, J.~E., {et~al.} 2021, Reviews of Modern Physics, 93, 015002

\bibitem[{{Cristallo} {et~al.}(2011){Cristallo}, {Piersanti}, {Straniero}, {Gallino}, {Dom{\'\i}nguez}, {Abia}, {Di Rico}, {Quintini}, \& {Bisterzo}}]{Cristallo2011ApJS}
{Cristallo}, S., {Piersanti}, L., {Straniero}, O., {et~al.} 2011, \apjs, 197, 17

\bibitem[{{Cristallo} {et~al.}(2015){Cristallo}, {Straniero}, {Piersanti}, \& {Gobrecht}}]{Cristallo2015ApJS}
{Cristallo}, S., {Straniero}, O., {Piersanti}, L., \& {Gobrecht}, D. 2015, \apjs, 219, 40

\bibitem[{{Cyburt} {et~al.}(2010){Cyburt}, {Amthor}, {Ferguson}, {Meisel}, {Smith}, {Warren}, {Heger}, {Hoffman}, {Rauscher}, {Sakharuk}, {Schatz}, {Thielemann}, \& {Wiescher}}]{Cyburt2010ApJS}
{Cyburt}, R.~H., {Amthor}, A.~M., {Ferguson}, R., {et~al.} 2010, \apjs, 189, 240

\bibitem[{Danielewicz \& Lee(2014)}]{Danielewicz2013NPA}
Danielewicz, P. \& Lee, J. 2014, Nucl. Phys. A, 922, 1

\bibitem[{{Domoto} {et~al.}(2022){Domoto}, {Tanaka}, {Kato}, {Kawaguchi}, {Hotokezaka}, \& {Wanajo}}]{Domoto2022ApJ}
{Domoto}, N., {Tanaka}, M., {Kato}, D., {et~al.} 2022, \apj, 939, 8

\bibitem[{{Domoto} {et~al.}(2021){Domoto}, {Tanaka}, {Wanajo}, \& {Kawaguchi}}]{Domoto2021ApJ}
{Domoto}, N., {Tanaka}, M., {Wanajo}, S., \& {Kawaguchi}, K. 2021, \apj, 913, 26

\bibitem[{{Domoto} {et~al.}(2025){Domoto}, {Wanajo}, {Tanaka}, {Kato}, \& {Hotokezaka}}]{Domoto2025ApJ}
{Domoto}, N., {Wanajo}, S., {Tanaka}, M., {Kato}, D., \& {Hotokezaka}, K. 2025, \apj, 978, 99

\bibitem[{Drischler {et~al.}(2019)Drischler, Hebeler, \& Schwenk}]{Drischler2017PRL}
Drischler, C., Hebeler, K., \& Schwenk, A. 2019, Phys. Rev. Lett., 122, 042501

\bibitem[{{Eichler} {et~al.}(2019){Eichler}, {Sayar}, {Arcones}, \& {Rauscher}}]{Eichler2019ApJ}
{Eichler}, M., {Sayar}, W., {Arcones}, A., \& {Rauscher}, T. 2019, \apj, 879, 47

\bibitem[{{Einfeldt}(1988)}]{Einfeldt1988SJNA}
{Einfeldt}, B. 1988, SIAM Journal on Numerical Analysis, 25, 294

\bibitem[{{Etienne} {et~al.}(2021){Etienne}, {Brandt}, {Diener}, {Gabella}, {Gracia-Linares}, {Haas}, {Kedia}, {Alcubierre}, {Alic}, {Allen}, {Ansorg}, {Babiuc-Hamilton}, {Baiotti}, {Benger}, {Bentivegna}, {Bernuzzi}, {Bode}, {Bozzola}, {Brendal}, {Bruegmann}, {Campanelli}, {Cipolletta}, {Corvino}, {Cupp}, {De Pietri}, {Dimmelmeier}, {Dooley}, {Dorband}, {Elley}, {El Khamra}, {Faber}, {Font}, {Frieben}, {Giacomazzo}, {Goodale}, {Gundlach}, {Hawke}, {Hawley}, {Hinder}, {Huerta}, {Husa}, {Iyer}, {Johnson}, {Kellermann}, {Knapp}, {Koppitz}, {Laguna}, {Lanferman}, {L{\"o}ffler}, {Masso}, {Menger}, {Merzky}, {Maxwell Miller}, {Miller}, {Moesta}, {Montero}, {Mundim}, {Nerozzi}, {Noble}, {Ott}, {Paruchuri}, {Pollney}, {Radice}, {Radke}, {Reisswig}, {Rezzolla}, {Rideout}, {Ripeanu}, {Sala}, {Schewtschenko}, {Schnetter}, {Schutz}, {Seidel}, {Seidel}, {Shalf}, {Sible}, {Sperhake}, {Stergioulas}, {Suen}, {Szilagyi}, {Takahashi}, {Thomas}, {Thornburg}, {Tobias}, {Tonita}, {Walker}, {Wan}, {Wardell}, {Werneck}, {Witek},
  {Zilh{\~a}o}, {Zink}, \& {Zlochower}}]{Etienne2021zndo}
{Etienne}, Z., {Brandt}, S.~R., {Diener}, P., {et~al.} 2021, {The Einstein Toolkit}

\bibitem[{{Fahlman} {et~al.}(2023){Fahlman}, {Fern{\'a}ndez}, \& {Morsink}}]{Fahlman2023MNRAS}
{Fahlman}, S., {Fern{\'a}ndez}, R., \& {Morsink}, S. 2023, \mnras, 526, 952

\bibitem[{{Foucart} {et~al.}(2016){Foucart}, {Haas}, {Duez}, {O'Connor}, {Ott}, {Roberts}, {Kidder}, {Lippuner}, {Pfeiffer}, \& {Scheel}}]{Foucart2016PRDa}
{Foucart}, F., {Haas}, R., {Duez}, M.~D., {et~al.} 2016, \prd, 93, 044019

\bibitem[{{Foucart} {et~al.}(2021){Foucart}, {M{\"o}sta}, {Ramirez}, {Wright}, {Darbha}, \& {Kasen}}]{Foucart2021PhRvD}
{Foucart}, F., {M{\"o}sta}, P., {Ramirez}, T., {et~al.} 2021, \prd, 104, 123010

\bibitem[{{Fujibayashi} {et~al.}(2018){Fujibayashi}, {Kiuchi}, {Nishimura}, {Sekiguchi}, \& {Shibata}}]{Fujibayashi2018ApJ}
{Fujibayashi}, S., {Kiuchi}, K., {Nishimura}, N., {Sekiguchi}, Y., \& {Shibata}, M. 2018, \apj, 860, 64

\bibitem[{{Fujibayashi} {et~al.}(2020){Fujibayashi}, {Wanajo}, {Kiuchi}, {Kyutoku}, {Sekiguchi}, \& {Shibata}}]{Fujibayashi2020ApJ}
{Fujibayashi}, S., {Wanajo}, S., {Kiuchi}, K., {et~al.} 2020, \apj, 901, 122

\bibitem[{{Galeazzi} {et~al.}(2013){Galeazzi}, {Kastaun}, {Rezzolla}, \& {Font}}]{Galeazzi2013PhRvD}
{Galeazzi}, F., {Kastaun}, W., {Rezzolla}, L., \& {Font}, J.~A. 2013, \prd, 88, 064009

\bibitem[{{Gillanders} {et~al.}(2024){Gillanders}, {Sim}, {Smartt}, {Goriely}, \& {Bauswein}}]{Gillanders2024MNRAS}
{Gillanders}, J.~H., {Sim}, S.~A., {Smartt}, S.~J., {Goriely}, S., \& {Bauswein}, A. 2024, \mnras, 529, 2918

\bibitem[{{Gourgoulhon} {et~al.}(2016){Gourgoulhon}, {Grandcl{\'e}ment}, {Marck}, {Novak}, \& {Taniguchi}}]{Gourgoulhon2016}
{Gourgoulhon}, E., {Grandcl{\'e}ment}, P., {Marck}, J.-A., {Novak}, J., \& {Taniguchi}, K. 2016, {LORENE: Spectral methods differential equations solver}

\bibitem[{{Gourgoulhon} {et~al.}(2001){Gourgoulhon}, {Grandcl{\'e}ment}, {Taniguchi}, {Marck}, \& {Bonazzola}}]{Gourgoulhon2001PhRvD}
{Gourgoulhon}, E., {Grandcl{\'e}ment}, P., {Taniguchi}, K., {Marck}, J.-A., \& {Bonazzola}, S. 2001, \prd, 63, 064029

\bibitem[{{Harten} {et~al.}(1983){Harten}, {Lax}, \& {Leer}}]{Harten1983SIAM}
{Harten}, A., {Lax}, P.~D., \& {Leer}, B.~v. 1983, SIAM Review, 25, 35

\bibitem[{{Hempel} \& {Schaffner-Bielich}(2010)}]{Hempel2010NPhysA}
{Hempel}, M. \& {Schaffner-Bielich}, J. 2010, \nphysa, 837, 210

\bibitem[{{Hilditch} {et~al.}(2013){Hilditch}, {Bernuzzi}, {Thierfelder}, {Cao}, {Tichy}, \& {Br{\"u}gmann}}]{Hilditch2013PrD}
{Hilditch}, D., {Bernuzzi}, S., {Thierfelder}, M., {et~al.} 2013, \prd, 88, 084057

\bibitem[{{Hoffman} {et~al.}(1997{\natexlab{a}}){Hoffman}, {Woosley}, \& {Qian}}]{Hoffman1997NuPhA}
{Hoffman}, R.~D., {Woosley}, S.~E., \& {Qian}, Y.-Z. 1997{\natexlab{a}}, \nphysa, 621, 397

\bibitem[{{Hoffman} {et~al.}(1997{\natexlab{b}}){Hoffman}, {Woosley}, \& {Qian}}]{Hoffman1997ApJ}
{Hoffman}, R.~D., {Woosley}, S.~E., \& {Qian}, Y.-Z. 1997{\natexlab{b}}, \apj, 482, 951

\bibitem[{{Holmbeck} {et~al.}(2018){Holmbeck}, {Beers}, {Roederer}, {Placco}, {Hansen}, {Sakari}, {Sneden}, {Liu}, {Lee}, {Cowan}, \& {Frebel}}]{Holmbeck2018ApJ}
{Holmbeck}, E.~M., {Beers}, T.~C., {Roederer}, I.~U., {et~al.} 2018, \apjl, 859, L24

\bibitem[{{Holmbeck} {et~al.}(2019){Holmbeck}, {Frebel}, {McLaughlin}, {Mumpower}, {Sprouse}, \& {Surman}}]{Holmbeck2019ApJ}
{Holmbeck}, E.~M., {Frebel}, A., {McLaughlin}, G.~C., {et~al.} 2019, \apj, 881, 5

\bibitem[{{Honda} {et~al.}(2006){Honda}, {Aoki}, {Ishimaru}, {Wanajo}, \& {Ryan}}]{Honda2006ApJ}
{Honda}, S., {Aoki}, W., {Ishimaru}, Y., {Wanajo}, S., \& {Ryan}, S.~G. 2006, \apj, 643, 1180

\bibitem[{{Horowitz}(2019)}]{Horowitz2019AnPhy}
{Horowitz}, C.~J. 2019, Annals of Physics, 411, 167992

\bibitem[{{Hotokezaka} {et~al.}(2023){Hotokezaka}, {Tanaka}, {Kato}, \& {Gaigalas}}]{Hotokezaka2023MNRAS}
{Hotokezaka}, K., {Tanaka}, M., {Kato}, D., \& {Gaigalas}, G. 2023, \mnras, 526, L155

\bibitem[{{Jacobi} {et~al.}(2024){Jacobi}, {Guercilena}, {Huth}, {Ricigliano}, {Arcones}, \& {Schwenk}}]{Jacobi2024MNRAS}
{Jacobi}, M., {Guercilena}, F.~M., {Huth}, S., {et~al.} 2024, \mnras, 527, 8812

\bibitem[{{Jacobi} {et~al.}(2026){Jacobi}, {Magistrelli}, {Loffredo}, {Ricigliano}, {Chiesa}, {Bernuzzi}, {Perego}, \& {Arcones}}]{Jacobi2026ApJ}
{Jacobi}, M., {Magistrelli}, F., {Loffredo}, E., {et~al.} 2026, \apjl, 999, L16

\bibitem[{{Ji} \& {Frebel}(2018)}]{Ji2018ApJ}
{Ji}, A.~P. \& {Frebel}, A. 2018, \apj, 856, 138

\bibitem[{{Just} {et~al.}(2015){Just}, {Obergaulinger}, \& {Janka}}]{Just2015MNRAS}
{Just}, O., {Obergaulinger}, M., \& {Janka}, H.-T. 2015, \mnras, 453, 3386

\bibitem[{{Just} {et~al.}(2023){Just}, {Vijayan}, {Xiong}, {Goriely}, {Soultanis}, {Bauswein}, {Guilet}, {Janka}, \& {Mart{\'\i}nez-Pinedo}}]{Just2023ApJ}
{Just}, O., {Vijayan}, V., {Xiong}, Z., {et~al.} 2023, \apjl, 951, L12

\bibitem[{{Kawaguchi} {et~al.}(2021){Kawaguchi}, {Fujibayashi}, {Shibata}, {Tanaka}, \& {Wanajo}}]{Kawaguchi2021ApJ}
{Kawaguchi}, K., {Fujibayashi}, S., {Shibata}, M., {Tanaka}, M., \& {Wanajo}, S. 2021, \apj, 913, 100

\bibitem[{{Khuyagbaatar}(2020)}]{Khuyagbaatar2020NuPhA100221958K}
{Khuyagbaatar}, J. 2020, \nphysa, 1002, 121958

\bibitem[{{Kiuchi} {et~al.}(2023){Kiuchi}, {Fujibayashi}, {Hayashi}, {Kyutoku}, {Sekiguchi}, \& {Shibata}}]{Kiuchi2023PhRvL}
{Kiuchi}, K., {Fujibayashi}, S., {Hayashi}, K., {et~al.} 2023, \prl, 131, 011401

\bibitem[{{Kodama} \& {Takahashi}(1975)}]{Kodama1975NuPhA.239..489K}
{Kodama}, T. \& {Takahashi}, K. 1975, \nphysa, 239, 489

\bibitem[{{Koning} {et~al.}(2023){Koning}, {Hilaire}, \& {Goriely}}]{Koning2023EPJA}
{Koning}, A., {Hilaire}, S., \& {Goriely}, S. 2023, European Physical Journal A, 59, 131

\bibitem[{{Kullmann} {et~al.}(2022){Kullmann}, {Goriely}, {Just}, {Ardevol-Pulpillo}, {Bauswein}, \& {Janka}}]{Kullman2022MNRAS}
{Kullmann}, I., {Goriely}, S., {Just}, O., {et~al.} 2022, \mnras, 510, 2804

\bibitem[{{Kullmann} {et~al.}(2023){Kullmann}, {Goriely}, {Just}, {Bauswein}, \& {Janka}}]{Kullmann2023MNRAS}
{Kullmann}, I., {Goriely}, S., {Just}, O., {Bauswein}, A., \& {Janka}, H.-T. 2023, \mnras, 523, 2551

\bibitem[{{Lattimer} \& {Steiner}(2014)}]{Lattimer2014APJ}
{Lattimer}, J.~M. \& {Steiner}, A.~W. 2014, \apj, 784, 123

\bibitem[{{L{\"o}ffler} {et~al.}(2012){L{\"o}ffler}, {Faber}, {Bentivegna}, {Bode}, {Diener}, {Haas}, {Hinder}, {Mundim}, {Ott}, {Schnetter}, {Allen}, {Campanelli}, \& {Laguna}}]{Loffler2012CQGra}
{L{\"o}ffler}, F., {Faber}, J., {Bentivegna}, E., {et~al.} 2012, Classical and Quantum Gravity, 29, 115001

\bibitem[{{Longo Micchi} {et~al.}(2026){Longo Micchi}, {Magistrelli}, {Jacobi}, \& {Bernuzzi}}]{LongoMicchi2026arXiv}
{Longo Micchi}, L.~F., {Magistrelli}, F., {Jacobi}, M., \& {Bernuzzi}, S. 2026, arXiv e-prints, arXiv:2608.10100

\bibitem[{{Magistrelli} {et~al.}(2026){Magistrelli}, {Bernuzzi}, {Perego}, {Jacobi}, \& {Fontes}}]{Magistrelli2026A&A}
{Magistrelli}, F., {Bernuzzi}, S., {Perego}, A., {Jacobi}, M., \& {Fontes}, C.~J. 2026, \aap, 708, A126

\bibitem[{{Martinet} \& {Goriely}(2025)}]{Martinet2025A&A}
{Martinet}, S. \& {Goriely}, S. 2025, \aap, 694, A180

\bibitem[{{Miller} {et~al.}(2019){Miller}, {Lamb}, {Dittmann}, {Bogdanov}, {Arzoumanian}, {Gendreau}, {Guillot}, {Harding}, {Ho}, {Lattimer}, {Ludlam}, {Mahmoodifar}, {Morsink}, {Ray}, {Strohmayer}, {Wood}, {Enoto}, {Foster}, {Okajima}, {Prigozhin}, \& {Soong}}]{Miller2019ApJ}
{Miller}, M.~C., {Lamb}, F.~K., {Dittmann}, A.~J., {et~al.} 2019, \apjl, 887, L24

\bibitem[{{Miller} {et~al.}(2021){Miller}, {Lamb}, {Dittmann}, {Bogdanov}, {Arzoumanian}, {Gendreau}, {Guillot}, {Ho}, {Lattimer}, {Loewenstein}, {Morsink}, {Ray}, {Wolff}, {Baker}, {Cazeau}, {Manthripragada}, {Markwardt}, {Okajima}, {Pollard}, {Cognard}, {Cromartie}, {Fonseca}, {Guillemot}, {Kerr}, {Parthasarathy}, {Pennucci}, {Ransom}, \& {Stairs}}]{Miller2021ApJL}
{Miller}, M.~C., {Lamb}, F.~K., {Dittmann}, A.~J., {et~al.} 2021, \apjl, 918, L28

\bibitem[{{M{\"o}ller} {et~al.}(2019){M{\"o}ller}, {Mumpower}, {Kawano}, \& {Myers}}]{Moller2019ADNDT.125....1M}
{M{\"o}ller}, P., {Mumpower}, M.~R., {Kawano}, T., \& {Myers}, W.~D. 2019, Atomic Data and Nuclear Data Tables, 125, 1

\bibitem[{{M{\"o}ller} {et~al.}(2015){M{\"o}ller}, {Sierk}, {Ichikawa}, {Iwamoto}, \& {Mumpower}}]{Moller2015PhRvC..91b4310M}
{M{\"o}ller}, P., {Sierk}, A.~J., {Ichikawa}, T., {Iwamoto}, A., \& {Mumpower}, M. 2015, \prc, 91, 024310

\bibitem[{{M{\"o}ller} {et~al.}(2016){M{\"o}ller}, {Sierk}, {Ichikawa}, \& {Sagawa}}]{Moller2016ADNDT}
{M{\"o}ller}, P., {Sierk}, A.~J., {Ichikawa}, T., \& {Sagawa}, H. 2016, Atomic Data and Nuclear Data Tables, 109, 1

\bibitem[{{Mumpower} {et~al.}(2020){Mumpower}, {Jaffke}, {Verriere}, \& {Randrup}}]{Mumpower2020PhRvC.101e4607M}
{Mumpower}, M.~R., {Jaffke}, P., {Verriere}, M., \& {Randrup}, J. 2020, \prc, 101, 054607

\bibitem[{{National Research Council}(2003)}]{NAP10079}
{National Research Council}. 2003, {Connecting Quarks with the Cosmos: Eleven Science Questions for the New Century} (Washington, DC: The National Academies Press)

\bibitem[{{Nedora} {et~al.}(2021){Nedora}, {Bernuzzi}, {Radice}, {Daszuta}, {Endrizzi}, {Perego}, {Prakash}, {Safarzadeh}, {Schianchi}, \& {Logoteta}}]{Nedora2021ApJ}
{Nedora}, V., {Bernuzzi}, S., {Radice}, D., {et~al.} 2021, \apj, 906, 98

\bibitem[{{Nishimura} {et~al.}(2015){Nishimura}, {Takiwaki}, \& {Thielemann}}]{Nishimura2015ApJ}
{Nishimura}, N., {Takiwaki}, T., \& {Thielemann}, F.-K. 2015, \apj, 810, 109

\bibitem[{{Oechslin} {et~al.}(2007){Oechslin}, {Janka}, \& {Marek}}]{Oechslin2007A&A}
{Oechslin}, R., {Janka}, H.-T., \& {Marek}, A. 2007, \aap, 467, 395

\bibitem[{{Oertel} {et~al.}(2017){Oertel}, {Hempel}, {Kl{\"a}hn}, \& {Typel}}]{Oertel2017RvMP}
{Oertel}, M., {Hempel}, M., {Kl{\"a}hn}, T., \& {Typel}, S. 2017, Reviews of Modern Physics, 89, 015007

\bibitem[{{Panov} {et~al.}(2005){Panov}, {Kolbe}, {Pfeiffer}, {Rauscher}, {Kratz}, \& {Thielemann}}]{Panov2005NuPhA.747..633P}
{Panov}, I.~V., {Kolbe}, E., {Pfeiffer}, B., {et~al.} 2005, \nphysa, 747, 633

\bibitem[{{Panov} {et~al.}(2010){Panov}, {Korneev}, {Rauscher}, {Mart{\'\i}nez-Pinedo}, {Keli{\'c}-Heil}, {Zinner}, \& {Thielemann}}]{Panov2010A&A...513A..61P}
{Panov}, I.~V., {Korneev}, I.~Y., {Rauscher}, T., {et~al.} 2010, \aap, 513, A61

\bibitem[{{Perego} {et~al.}(2019){Perego}, {Bernuzzi}, \& {Radice}}]{Perego2019EPJA}
{Perego}, A., {Bernuzzi}, S., \& {Radice}, D. 2019, European Physical Journal A, 55, 124

\bibitem[{{Perego} {et~al.}(2021){Perego}, {Thielemann}, \& {Cescutti}}]{Perego2021hgwa}
{Perego}, A., {Thielemann}, F.~K., \& {Cescutti}, G. 2021, in Handbook of Gravitational Wave Astronomy, 13

\bibitem[{{Perego} {et~al.}(2022){Perego}, {Vescovi}, {Fiore}, {Chiesa}, {Vogl}, {Benetti}, {Bernuzzi}, {Branchesi}, {Cappellaro}, {Cristallo}, {Fl{\"o}rs}, {Kerzendorf}, \& {Radice}}]{Perego2022ApJ}
{Perego}, A., {Vescovi}, D., {Fiore}, A., {et~al.} 2022, \apj, 925, 22

\bibitem[{{Pian} {et~al.}(2017){Pian}, {D'Avanzo}, {Benetti}, {Branchesi}, {Brocato}, {Campana}, {Cappellaro}, {Covino}, {D'Elia}, {Fynbo}, {Getman}, {Ghirlanda}, {Ghisellini}, {Grado}, {Greco}, {Hjorth}, {Kouveliotou}, {Levan}, {Limatola}, {Malesani}, {Mazzali}, {Melandri}, {M{\o}ller}, {Nicastro}, {Palazzi}, {Piranomonte}, {Rossi}, {Salafia}, {Selsing}, {Stratta}, {Tanaka}, {Tanvir}, {Tomasella}, {Watson}, {Yang}, {Amati}, {Antonelli}, {Ascenzi}, {Bernardini}, {Bo{\"e}r}, {Bufano}, {Bulgarelli}, {Capaccioli}, {Casella}, {Castro-Tirado}, {Chassande-Mottin}, {Ciolfi}, {Copperwheat}, {Dadina}, {De Cesare}, {di Paola}, {Fan}, {Gendre}, {Giuffrida}, {Giunta}, {Hunt}, {Israel}, {Jin}, {Kasliwal}, {Klose}, {Lisi}, {Longo}, {Maiorano}, {Mapelli}, {Masetti}, {Nava}, {Patricelli}, {Perley}, {Pescalli}, {Piran}, {Possenti}, {Pulone}, {Razzano}, {Salvaterra}, {Schipani}, {Spera}, {Stamerra}, {Stella}, {Tagliaferri}, {Testa}, {Troja}, {Turatto}, {Vergani}, \& {Vergani}}]{Pian2017Natur}
{Pian}, E., {D'Avanzo}, P., {Benetti}, S., {et~al.} 2017, \nat, 551, 67

\bibitem[{{Prantzos} {et~al.}(2020){Prantzos}, {Abia}, {Cristallo}, {Limongi}, \& {Chieffi}}]{Prantzos2020MNRAS}
{Prantzos}, N., {Abia}, C., {Cristallo}, S., {Limongi}, M., \& {Chieffi}, A. 2020, \mnras, 491, 1832

\bibitem[{{Raaijmakers} {et~al.}(2021){Raaijmakers}, {Greif}, {Hebeler}, {Hinderer}, {Nissanke}, {Schwenk}, {Riley}, {Watts}, {Lattimer}, \& {Ho}}]{Raaijmakers2021APJL}
{Raaijmakers}, G., {Greif}, S.~K., {Hebeler}, K., {et~al.} 2021, \apjl, 918, L29

\bibitem[{{Radice} \& {Bernuzzi}(2023)}]{Radice2023ApJ}
{Radice}, D. \& {Bernuzzi}, S. 2023, \apj, 959, 46

\bibitem[{{Radice} \& {Bernuzzi}(2024)}]{Radice2024JPhCS}
{Radice}, D. \& {Bernuzzi}, S. 2024, in Journal of Physics Conference Series, Vol. 2742, Journal of Physics Conference Series (IOP), 012009

\bibitem[{{Radice} {et~al.}(2020){Radice}, {Bernuzzi}, \& {Perego}}]{Radice2020ARNPS}
{Radice}, D., {Bernuzzi}, S., \& {Perego}, A. 2020, Annual Review of Nuclear and Particle Science, 70, 95

\bibitem[{{Radice} {et~al.}(2022){Radice}, {Bernuzzi}, {Perego}, \& {Haas}}]{Radice2022MNRAS}
{Radice}, D., {Bernuzzi}, S., {Perego}, A., \& {Haas}, R. 2022, \mnras, 512, 1499

\bibitem[{{Radice} {et~al.}(2018){Radice}, {Perego}, {Hotokezaka}, {Fromm}, {Bernuzzi}, \& {Roberts}}]{Radice2018ApJ}
{Radice}, D., {Perego}, A., {Hotokezaka}, K., {et~al.} 2018, \apj, 869, 130

\bibitem[{{Radice} \& {Rezzolla}(2012)}]{Radice2012AAP}
{Radice}, D. \& {Rezzolla}, L. 2012, \aap, 547, A26

\bibitem[{{Radice} {et~al.}(2014{\natexlab{a}}){Radice}, {Rezzolla}, \& {Galeazzi}}]{Radice2014APS}
{Radice}, D., {Rezzolla}, L., \& {Galeazzi}, F. 2014{\natexlab{a}}, in APS Meeting Abstracts, Vol. 2014, APS April Meeting Abstracts, R15.003

\bibitem[{{Radice} {et~al.}(2014{\natexlab{b}}){Radice}, {Rezzolla}, \& {Galeazzi}}]{Radice2014MNRAS}
{Radice}, D., {Rezzolla}, L., \& {Galeazzi}, F. 2014{\natexlab{b}}, \mnras, 437, L46

\bibitem[{{Radice} {et~al.}(2014{\natexlab{c}}){Radice}, {Rezzolla}, \& {Galeazzi}}]{Radice2014CQG}
{Radice}, D., {Rezzolla}, L., \& {Galeazzi}, F. 2014{\natexlab{c}}, Classical and Quantum Gravity, 31, 075012

\bibitem[{{Rahmouni} {et~al.}(2025){Rahmouni}, {Tanaka}, {Domoto}, {Kato}, {Hotokezaka}, {Aoki}, {Hirano}, {Kotani}, {Kuzuhara}, \& {Tamura}}]{Rahmouni2025ApJ}
{Rahmouni}, S., {Tanaka}, M., {Domoto}, N., {et~al.} 2025, \apj, 980, 43

\bibitem[{{Reichert} {et~al.}(2023){Reichert}, {Winteler}, {Korobkin}, {Arcones}, {Bliss}, {Eichler}, {Frischknecht}, {Fr{\"o}hlich}, {Hirschi}, {Jacobi}, {Kuske}, {Mart{\'\i}nez-Pinedo}, {Martin}, {Mocelj}, {Rauscher}, \& {Thielemann}}]{Reichert2023ApJS}
{Reichert}, M., {Winteler}, C., {Korobkin}, O., {et~al.} 2023, \apjs, 268, 66

\bibitem[{{Reisswig} {et~al.}(2013{\natexlab{a}}){Reisswig}, {Haas}, {Ott}, {Abdikamalov}, {M{\"o}sta}, {Pollney}, \& {Schnetter}}]{Reisswig2013PhRvD}
{Reisswig}, C., {Haas}, R., {Ott}, C.~D., {et~al.} 2013{\natexlab{a}}, \prd, 87, 064023

\bibitem[{{Reisswig} {et~al.}(2013{\natexlab{b}}){Reisswig}, {Ott}, {Abdikamalov}, {Haas}, {M{\"o}sta}, \& {Schnetter}}]{Reisswig2013PhRvL}
{Reisswig}, C., {Ott}, C.~D., {Abdikamalov}, E., {et~al.} 2013{\natexlab{b}}, \prl, 111, 151101

\bibitem[{{Ricigliano} {et~al.}(2024){Ricigliano}, {Jacobi}, \& {Arcones}}]{Ricigliano2024MNRAS2}
{Ricigliano}, G., {Jacobi}, M., \& {Arcones}, A. 2024, \mnras, 533, 2096

\bibitem[{{Riley} {et~al.}(2019){Riley}, {Watts}, {Bogdanov}, {Ray}, {Ludlam}, {Guillot}, {Arzoumanian}, {Baker}, {Bilous}, {Chakrabarty}, {Gendreau}, {Harding}, {Ho}, {Lattimer}, {Morsink}, \& {Strohmayer}}]{Riley2019ApJ}
{Riley}, T.~E., {Watts}, A.~L., {Bogdanov}, S., {et~al.} 2019, \apjl, 887, L21

\bibitem[{{Rochman} {et~al.}(2025){Rochman}, {Koning}, {Goriely}, \& {Hilaire}}]{Rochman2025NuPhA105322951R}
{Rochman}, D., {Koning}, A., {Goriely}, S., \& {Hilaire}, S. 2025, \nphysa, 1053, 122951

\bibitem[{{Roederer} {et~al.}(2009){Roederer}, {Kratz}, {Frebel}, {Christlieb}, {Pfeiffer}, {Cowan}, \& {Sneden}}]{Roederer2009ApJ}
{Roederer}, I.~U., {Kratz}, K.-L., {Frebel}, A., {et~al.} 2009, \apj, 698, 1963

\bibitem[{{Roederer} {et~al.}(2012){Roederer}, {Lawler}, {Sobeck}, {Beers}, {Cowan}, {Frebel}, {Ivans}, {Schatz}, {Sneden}, \& {Thompson}}]{Roederer2012ApJS}
{Roederer}, I.~U., {Lawler}, J.~E., {Sobeck}, J.~S., {et~al.} 2012, \apjs, 203, 27

\bibitem[{{Rosswog} {et~al.}(2025){Rosswog}, {Sarin}, {Nakar}, \& {Diener}}]{Rosswog2025MNRAS}
{Rosswog}, S., {Sarin}, N., {Nakar}, E., \& {Diener}, P. 2025, \mnras, 538, 907

\bibitem[{{Rosswog} {et~al.}(2018){Rosswog}, {Sollerman}, {Feindt}, {Goobar}, {Korobkin}, {Wollaeger}, {Fremling}, \& {Kasliwal}}]{Rosswog2018A&A}
{Rosswog}, S., {Sollerman}, J., {Feindt}, U., {et~al.} 2018, \aap, 615, A132

\bibitem[{{Schnetter} {et~al.}(2004){Schnetter}, {Hawley}, \& {Hawke}}]{Schnetter2004CQGra}
{Schnetter}, E., {Hawley}, S.~H., \& {Hawke}, I. 2004, Classical and Quantum Gravity, 21, 1465

\bibitem[{{Shibata} \& {Hotokezaka}(2019)}]{Shibata2019ARNPS}
{Shibata}, M. \& {Hotokezaka}, K. 2019, Annual Review of Nuclear and Particle Science, 69, 41

\bibitem[{{Shlomo} {et~al.}(2006){Shlomo}, {Kolomietz}, \& {Col{\`o}}}]{Shlomo2006EPJA}
{Shlomo}, S., {Kolomietz}, V.~M., \& {Col{\`o}}, G. 2006, European Physical Journal A, 30, 23

\bibitem[{{Siegel} \& {Metzger}(2017)}]{Siegel2017PhRvL}
{Siegel}, D.~M. \& {Metzger}, B.~D. 2017, \prl, 119, 231102

\bibitem[{{Smartt} {et~al.}(2017){Smartt}, {Chen}, {Jerkstrand}, {Coughlin}, {Kankare}, {Sim}, {Fraser}, {Inserra}, {Maguire}, {Chambers}, {Huber}, {Kr{\"u}hler}, {Leloudas}, {Magee}, {Shingles}, {Smith}, {Young}, {Tonry}, {Kotak}, {Gal-Yam}, {Lyman}, {Homan}, {Agliozzo}, {Anderson}, {Angus}, {Ashall}, {Barbarino}, {Bauer}, {Berton}, {Botticella}, {Bulla}, {Bulger}, {Cannizzaro}, {Cano}, {Cartier}, {Cikota}, {Clark}, {De Cia}, {Della Valle}, {Denneau}, {Dennefeld}, {Dessart}, {Dimitriadis}, {Elias-Rosa}, {Firth}, {Flewelling}, {Fl{\"o}rs}, {Franckowiak}, {Frohmaier}, {Galbany}, {Gonz{\'a}lez-Gait{\'a}n}, {Greiner}, {Gromadzki}, {Guelbenzu}, {Guti{\'e}rrez}, {Hamanowicz}, {Hanlon}, {Harmanen}, {Heintz}, {Heinze}, {Hernandez}, {Hodgkin}, {Hook}, {Izzo}, {James}, {Jonker}, {Kerzendorf}, {Klose}, {Kostrzewa-Rutkowska}, {Kowalski}, {Kromer}, {Kuncarayakti}, {Lawrence}, {Lowe}, {Magnier}, {Manulis}, {Martin-Carrillo}, {Mattila}, {McBrien}, {M{\"u}ller}, {Nordin}, {O'Neill}, {Onori}, {Palmerio}, {Pastorello},
  {Patat}, {Pignata}, {Podsiadlowski}, {Pumo}, {Prentice}, {Rau}, {Razza}, {Rest}, {Reynolds}, {Roy}, {Ruiter}, {Rybicki}, {Salmon}, {Schady}, {Schultz}, {Schweyer}, {Seitenzahl}, {Smith}, {Sollerman}, {Stalder}, {Stubbs}, {Sullivan}, {Szegedi}, {Taddia}, {Taubenberger}, {Terreran}, {van Soelen}, {Vos}, {Wainscoat}, {Walton}, {Waters}, {Weiland}, {Willman}, {Wiseman}, {Wright}, {Wyrzykowski}, \& {Yaron}}]{Smartt2017Natur}
{Smartt}, S.~J., {Chen}, T.~W., {Jerkstrand}, A., {et~al.} 2017, \nat, 551, 75

\bibitem[{{Sneden} {et~al.}(2008){Sneden}, {Cowan}, \& {Gallino}}]{Sneden2008ARA&A}
{Sneden}, C., {Cowan}, J.~J., \& {Gallino}, R. 2008, \araa, 46, 241

\bibitem[{{Sneppen} {et~al.}(2024){Sneppen}, {Damgaard}, {Watson}, {Collins}, {Shingles}, \& {Sim}}]{Sneppen2024A&A}
{Sneppen}, A., {Damgaard}, R., {Watson}, D., {et~al.} 2024, \aap, 692, A134

\bibitem[{{Sneppen} {et~al.}(2026){Sneppen}, {Just}, {Bauswein}, {Damgaard}, {Watson}, {Shingles}, {Collins}, {Sim}, {Xiong}, {Mart{\'\i}nez-Pinedo}, {Soultanis}, \& {Vijayan}}]{Sneppen2026PhRvD}
{Sneppen}, A., {Just}, O., {Bauswein}, A., {et~al.} 2026, \prd, 113, 063038

\bibitem[{{Sneppen} \& {Watson}(2023)}]{Sneppen2023A&A}
{Sneppen}, A. \& {Watson}, D. 2023, \aap, 675, A194

\bibitem[{Sprouse {et~al.}(2024)Sprouse, Lund, Miller, McLaughlin, \& Mumpower}]{Sprouse2023ApJ}
Sprouse, T.~M., Lund, K.~A., Miller, J.~M., McLaughlin, G.~C., \& Mumpower, M.~R. 2024, Astrophys. J., 962, 79

\bibitem[{{Steiner} {et~al.}(2005){Steiner}, {Prakash}, {Lattimer}, \& {Ellis}}]{Steiner2005PhysRep}
{Steiner}, A.~W., {Prakash}, M., {Lattimer}, J.~M., \& {Ellis}, P.~J. 2005, \physrep, 411, 325

\bibitem[{{Suresh} \& {Huynh}(1997)}]{Suresh1997JCoPh}
{Suresh}, A. \& {Huynh}, H.~T. 1997, Journal of Computational Physics, 136, 83

\bibitem[{{Thielemann} {et~al.}(2017){Thielemann}, {Eichler}, {Panov}, \& {Wehmeyer}}]{Thielemann2017ARNPS}
{Thielemann}, F.~K., {Eichler}, M., {Panov}, I.~V., \& {Wehmeyer}, B. 2017, Annual Review of Nuclear and Particle Science, 67, 253

\bibitem[{{Typel} {et~al.}(2010){Typel}, {R{\"o}pke}, {Kl{\"a}hn}, {Blaschke}, \& {Wolter}}]{Typel2010PRC}
{Typel}, S., {R{\"o}pke}, G., {Kl{\"a}hn}, T., {Blaschke}, D., \& {Wolter}, H.~H. 2010, \prc, 81, 015803

\bibitem[{{Vincent} {et~al.}(2020){Vincent}, {Foucart}, {Duez}, {Haas}, {Kidder}, {Pfeiffer}, \& {Scheel}}]{Vincent2020PhRvD}
{Vincent}, T., {Foucart}, F., {Duez}, M.~D., {et~al.} 2020, \prd, 101, 044053

\bibitem[{{Wanajo} \& {Ishimaru}(2006)}]{Wanajo2006NuPhA}
{Wanajo}, S. \& {Ishimaru}, Y. 2006, \nphysa, 777, 676

\bibitem[{{Wang} {et~al.}(2021){Wang}, {Huang}, {Kondev}, {Audi}, \& {Naimi}}]{Wang2021ChPhC..45c0003W}
{Wang}, M., {Huang}, W.~J., {Kondev}, F.~G., {Audi}, G., \& {Naimi}, S. 2021, Chinese Physics C, 45, 030003

\bibitem[{{Watson} {et~al.}(2019){Watson}, {Hansen}, {Selsing}, {Koch}, {Malesani}, {Andersen}, {Fynbo}, {Arcones}, {Bauswein}, {Covino}, {Grado}, {Heintz}, {Hunt}, {Kouveliotou}, {Leloudas}, {Levan}, {Mazzali}, \& {Pian}}]{Watson2019Nature}
{Watson}, D., {Hansen}, C.~J., {Selsing}, J., {et~al.} 2019, \nat, 574, 497

\bibitem[{{Witt} {et~al.}(2021){Witt}, {Psaltis}, {Yasin}, {Horn}, {Reichert}, {Kuroda}, {Obergaulinger}, {Couch}, \& {Arcones}}]{Witt2021ApJ}
{Witt}, M., {Psaltis}, A., {Yasin}, H., {et~al.} 2021, \apj, 921, 19

\bibitem[{{Xiong} {et~al.}(2024){Xiong}, {Mart{\'\i}nez-Pinedo}, {Just}, \& {Sieverding}}]{Xiong2024PhRvL}
{Xiong}, Z., {Mart{\'\i}nez-Pinedo}, G., {Just}, O., \& {Sieverding}, A. 2024, \prl, 132, 192701

\bibitem[{{Zappa} {et~al.}(2023){Zappa}, {Bernuzzi}, {Radice}, \& {Perego}}]{Zappa2023MNRAS}
{Zappa}, F., {Bernuzzi}, S., {Radice}, D., \& {Perego}, A. 2023, \mnras, 520, 1481

\end{thebibliography}

\begin{appendix}

\section{Testing tracers mass assignment}

\begin{figure*}[]
    \centering
    \includegraphics[scale = 0.45]{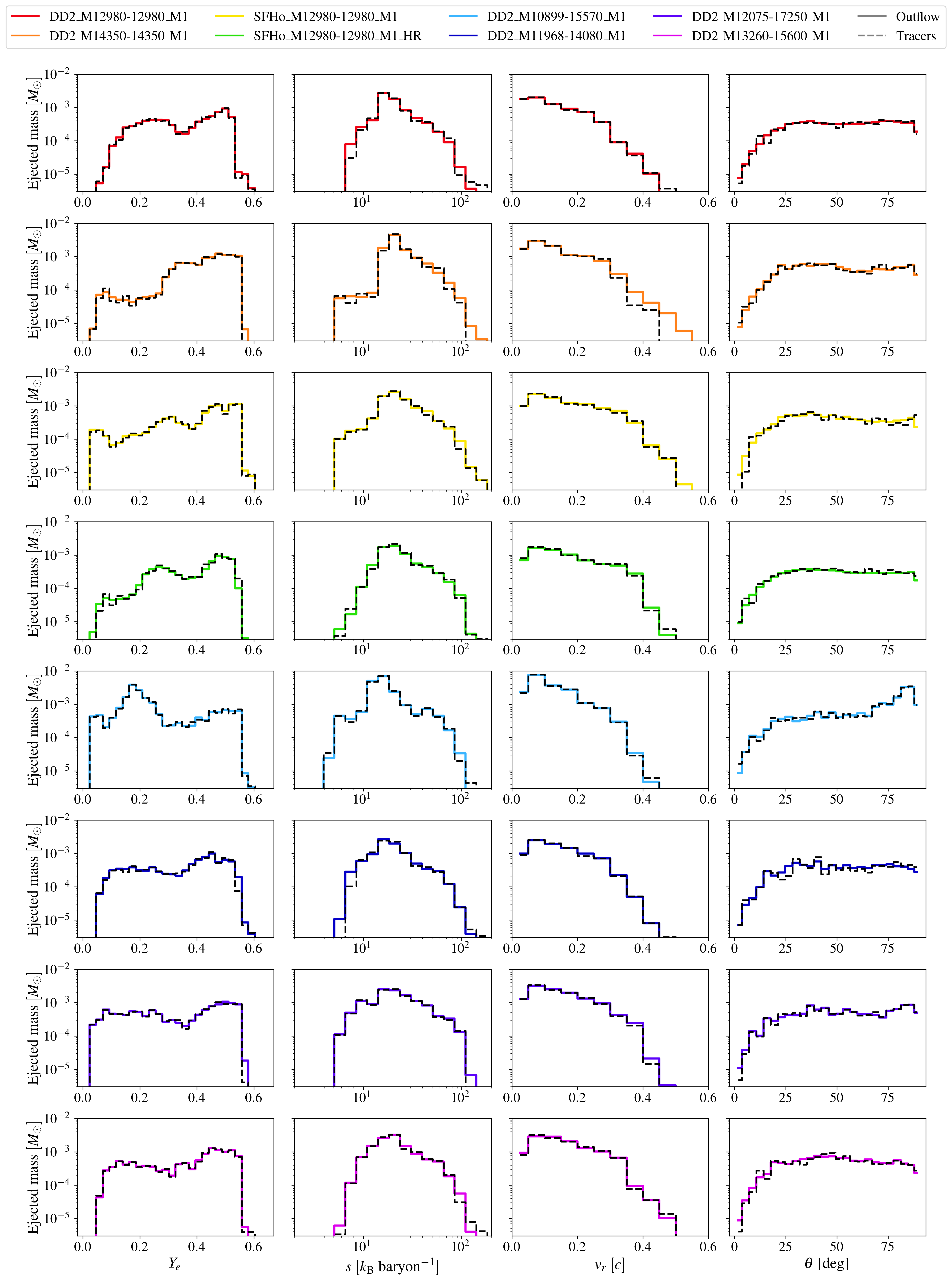}
    \caption{Mass distribution computed from outflow (solid lines) and mass-weighted tracers (dashed black lines) as function of $Y_e$, $s$, $v_r$, and $\theta$ for all simulations reported in \reftab{tab:ejecta}.}
    \label{fig:mass_distr_all_check}
\end{figure*}

We tested the tracer mass assignment procedure described in \refsec{subsec:trcrs_part} by comparing, for each simulation, the mass distributions as functions of the electron fraction $Y_e$, entropy $s$, radial velocity $v_r$, and polar angle $\theta$ obtained directly from the mass outflow with those reconstructed from the mass-weighted tracers. The comparison was performed for every ejection time window considered in our analysis. We find an extremely good agreement between the mass distributions computed with the two different methods. In \reffig{fig:mass_distr_all_check} we show the comparison for the total ejecta, where the mass distributions reconstructed from the tracers (black dashed lines) closely overlap those computed from the outflow data (solid lines), demonstrating that the adopted mass assignment accurately reproduces the properties of the ejecta. 

\end{appendix}

\end{document}